\documentclass[11pt]{article}%
\usepackage[pdftex]{graphicx}
\usepackage{verbatim,color}
\usepackage{amsmath}
\usepackage{subfigure}
\usepackage{setspace}
\usepackage{subfig}
\usepackage{url}
\usepackage{float}

\usepackage[left=1in,top=1in,bottom=1in,right=1in, nohead]%
{geometry}
\usepackage{amsfonts}
\usepackage{amssymb}
\usepackage{graphicx}%
\graphicspath{{./figures/}}  

\begin{document}

\title{Co-movements in financial fluctuations are anchored to economic fundamentals: A mesoscopic mapping}

\author{
Kiran Sharma\thanks{School of Computational and Integrative Sciences, Jawaharlal Nehru University, New Delhi-110067, India. Email: kiran34\_sit@jnu.ac.in}
\and Balagopal Gopalakrishnan\thanks{Finance and Accounting area, Indian Institute of Management, Vastrapur, 
Ahmedabad-380015, India. Email: balagopalg@iima.ac.in} 
\and
Anindya S. Chakrabarti\thanks {Economics area, Indian Institute of Management, Vastrapur, Ahmedabad-380015, India. Email: anindyac@iima.ac.in.} 
\and Anirban Chakraborti \thanks{School of Computational and Integrative Sciences, Jawaharlal Nehru University, New Delhi-110067, India. Email: anirban@jnu.ac.in}
}

\maketitle

\begin{abstract}
We demonstrate the existence of an empirical linkage between the
nominal financial networks and the underlying economic fundamentals across countries. 
We construct the nominal return correlation networks from daily data to encapsulate sector-level dynamics
and figure the relative importance of the sectors in the nominal network through a measure of centrality and clustering algorithms.
The eigenvector centrality  robustly identifies the backbone of the minimum spanning tree defined on the return networks as well as the primary cluster in the multidimensional scaling map.
We show that the sectors that are relatively large in size, defined with the metrics market capitalization, revenue and number of employees, constitute the core of the return networks, whereas the periphery is mostly populated by relatively smaller sectors. Therefore, sector-level nominal return dynamics is anchored to the real size effect, which ultimately shapes the optimal portfolios for risk management. Our results are reasonably robust across 27 countries of varying degrees of prosperity and across periods of market turbulence (2008-09) as well as relative calmness (2015-16).
\end{abstract}

\section{Introduction}
\label{intro}

Widespread existence of bubbles in the financial markets and extreme movements of return series indicate that the relationship between the macroeconomic fundamentals
and the asset prices is unstable \cite{Shiller_AER_81}. The `excess volatility puzzle' in the stock markets refers precisely to this disconnect between the volatility of asset returns
and the movements of the underlying fundamentals \cite{Gabaix_12}. 
Recent research emphasizes the roles played by wrong expectation, bounded rationality, herding behavior, etc. as being important causal factors for the observed disconnect \cite{Sornette_04}.
In this paper, we present an alternate view that the co-movements in financial assets are anchored to the corresponding macroeconomic fundamentals. 
Thus, nominal returns from individual assets might drift far from what can be predicted using expected cash-flow, while the joint evolution of the co-movement of returns 
are still related to aggregate size variables like market capitalization, revenue or number of employees.

In the following, we consider the economy to be a multi-layered network \cite{newman2010networks} defined over nodes at different levels of granularity, each having significantly different properties. 
At the micro level, firm size distributions show power law decays \cite{Axtell_01} and bi-exponential growth size distributions \cite{Stanley_96}. A scaling relationship between size of the firms and the corresponding volatiltity was also proposed \cite{Stanley_96}. 
At the macro level, 
similar features are seen, for example, as has been proposed  by \cite{Gabaix_11}. These suggest that there might be universal features of growth processes of economic entities (see also \cite{Stanley_98}).
Ref. \cite{Gabaix_11} also argued that the dispersion in relative sizes of firms contribute substantially to the aggregate volatility of an economy, providing a link from the micro level to the macro level.
A complementary view has emerged from the network literature that the dynamics at the intermediate sectoral level could play an important role in shaping the aggregate macro-level dynamics \cite{Acemoglu_12}.
We focus precisely on the `mesoscopic' level, which identifies with the production process of the economy while being granular enough to capture the network structure of co-movements in return fluctuations.

There are two modes of connectedness across sectors. At the nominal pricing level, the fluctuations of returns from the sectoral indices show the degree of co-movements across sectors. At the production level, the flow of goods and services across sectors \cite{Atalaya_11} gives rise to dispersion in relative sizes of these sectors. Here, we show that there exists a universal mapping between the inter-sectoral return dynamics and relative sizes of the sectors defined with multiple metrics, thus highlighting an empirical link between financial networks and macroscopic variables in a granular economy.
In particular, we show that the sectors with disproportionate shares of the economy, constitute the core of the corresponding return networks. 
Therefore, at the `mesoscopic' level, the dispersion in size explains the dispersion in `centralities' of nominal fluctuations of sectors.

To study the topology of the return network, we construct return correlation matrices from sectoral indices for 27 countries, and apply two commonly used clustering algorithms (minimum spanning tree and multi-dimensional scaling) to group sectors based on their co-movements. The influence of the sectors in the whole network can be found by using the eigenvector centrality, which is able to handle both directed as well as weighted graphs \cite{Jackson_10}. In this paper, we also propose a methodology to find a binary characterization of the `core-periphery' structure by using a modification of the eigenvector centrality. Such classification of the sectors according to whether they belong to the core or the periphery, allows one to pin down exactly which sectors are driving the
market correlations. We show that these sectors identified as core by the centrality measure, also constitute the backbone of the minimum spanning tree (MST) and cluster very closely in multi-dimensional scaling (MDS) maps, thereby confirming the robustness of our method of extraction of the core-periphery structure. 


To study the connection between the financial network with the underlying production process, we regress the eigenvector centrality measure on sector sizes defined with three different metrics, viz., market capitalization, revenue and employment, all aggregated at the sectoral level. The results across 27 countries clearly indicate that the dispersion in the economic size explains the variation in the dispersion of sectoral centralities in the correlation matrix. This is the primary finding of our paper, as it establishes the
the linkage between the economic fundamentals and the fluctuations of the return series. 
Finally, we study the risk diversification of a portfolio comprising sectoral indices, based on the eigenvector centralities. For the sake of simplicity, we use a rudimentary Markowitz portfolio allocation problem and show that the core sectors, i.e., the ones with sufficiently high centralities, do not usually appear in a minimum variance portfolio. Intuitively, very large sectors contribute significantly to the movement of the return correlations and they constitute the `market factor' of correlations. Hence, for reduction of the volatility of the portfolio, the weights assigned to such sectors contributing to the aggregate risk, are necessarily minimized.

We perform statistical tests on a comprehensive list of 27 countries that includes developed as well as developing countries across five continents, totaling 72 sectors in the financial economies. We base most of our studies on a recent and relatively calm period (2015-16), and then compare and contrast with a volatile period (2008-09), in order to check robustness of our findings across time. We show that the 2015-16 period gives very consistent results (25 out of 27 countries are in expected direction), whereas 2008-09 period is largely consistent (22 out of 26), although there are some aberrations as the number of statistically insignificant relationships increases. A consistent pathogenic case is Greece, which has been known to possess weak economic fundamentals along with severe crises in the financial markets in the recent times. 

\section{Data, Definitions and Methods}
\label{Sec:Materials}


\subsection{Data Description}
\label{subsec:data}
We have used the sectoral price indices from the Thomson Reuters Eikon database \cite{Thompson_reuters}, within the time frames January 2008- December 2009, and October 2014- September 2016. 
We have analyzed the data for a total of 72 sectors (see table \ref{Table:sectoral_index}), for the following countries: 
(1) \textbf{AUS}- Australia 
(2) \textbf{BEL}- Belgium 
(3) \textbf{CAN}- Canada 
(4) \textbf{CHE}- Switzerland 
(5) \textbf{DEU}- Germany 
(6) \textbf{DNK}- Denmark 
(7) \textbf{ESP}- Spain 
(8) \textbf{FIN}- Finland 
(9) \textbf{FRA}- France 
(10) \textbf{GBR}- United Kingdom
(11) \textbf{GRC}- Greece 
(12) \textbf{HKG}- Hong Kong 
(13) \textbf{IDN}- Indonesia
(14) \textbf{IND}- India  
(15) \textbf{JPN}- Japan 
(16) \textbf{LKA}- Sri Lanka 
(17) \textbf{MYS}- Malaysia 
(18) \textbf{NLD}- the Netherlands 
(19) \textbf{NOR}- Norway 
(20) \textbf{PHL}- Philippines 
(21) \textbf{PRT}- Portugal 
(22) \textbf{QAT}- Qatar 
(23) \textbf{SAU}- Saudi Arabia 
(24) \textbf{SWE}- Sweden
(25) \textbf{THA}- Thailand
(26) \textbf{USA}- United States of America
and
(27) \textbf{ZAF}- South Africa, 
spread across the continents of the Americas, Europe, Africa, Asia and Australia. 
The time series data on the real variables, such as market capitalization, revenue and the number of employees within each sector, are also available in the same database although at the company level rather than at the sectoral level. Hence, for our purposes of constructing sector-level macro aggregate variables, we collected the companies listed within each sector for one particular country, and then aggregated the relevant company-specific variables across all such companies within the corresponding sector.
 
We find that the USA economy is a good representative of the empirical results and hence, in the main text, we present the results for the USA economy in details. For the other 26 countries, the detailed results are presented in the Supplementary material. Note that data for Finland (FIN) was not available for the period 2008-09.


\begin{table}[H]
\centering
\caption{Abbreviations of the 72 sectors analyzed.}
\begin{tiny}
\begin{tabular}{|l|l|l|l|}
\hline
\textbf{Label} & \textbf{Sector}           & \textbf{Label} & \textbf{Sector}           \\ \hline
\textbf{AF}    & Agro \& Food Industry     & \textbf{MD}    & Media                     \\ \hline
\textbf{AG}    & Agriculture               & \textbf{MF}    & Manufacturing             \\ \hline
\textbf{AM}    & Automobiles               & \textbf{MG}    & Mining                    \\ \hline
\textbf{BC}    & Building \& Construction  & \textbf{MI}    & Multi Investments         \\ \hline
\textbf{BF}    & Banks \& Finance          & \textbf{MID}   & Miscellaneous Industries  \\ \hline
\textbf{BFT}   & Beverage, Food \& Tobacco & \textbf{MM}    & Metals \& Mining           \\ \hline
\textbf{BK}    & Bank                      & \textbf{MO}    & Mining \& Oil             \\ \hline
\textbf{BM}    & Basic Materials           & \textbf{MOT}   & Motors                    \\ \hline
\textbf{BR}    & Basic Resources           & \textbf{MP}    & Metal Products            \\ \hline
\textbf{CC}    & Consumer \& Cyclical      & \textbf{MP1}    & Media \& Publishing       \\ \hline
\textbf{CD}    & Consumer Discretionary    & \textbf{MT}    & Media \& Telecomm         \\ \hline
\textbf{CD1}   & Consumer Durables         & \textbf{OC}    & Oil \& Coal Products      \\ \hline
\textbf{CE}    & Cement                    & \textbf{OG}    & Oil and Gas               \\ \hline
\textbf{CG}    & Consumer Goods            & \textbf{PC}    & Property \& Construction  \\ \hline
\textbf{CG1}   & Capital Goods             & \textbf{PE}    & Power \& Energy           \\ \hline
\textbf{IT}    & Information  Technology   & \textbf{PG}    & Personal Goods            \\ \hline
\textbf{CH}    & Chemicals                 & \textbf{PH}    & PetroChemicals            \\ \hline
\textbf{CM}    & Consturction \& Materials & \textbf{PL}    & Plantation                \\ \hline
\textbf{CN}    & Construction              & \textbf{PR}    & Property                  \\ \hline
\textbf{CP}    & Consumer Products         & \textbf{PSU}   & Public Sector Undertaking \\ \hline
\textbf{CS}    & Consumer Staples          & \textbf{RB}    & Rubber                    \\ \hline
\textbf{CSR}   & Consumer Services         & \textbf{RE}    & Real Estate               \\ \hline
\textbf{EC}    & Energy \& Chemical        & \textbf{RT}    & Retail                    \\ \hline
\textbf{EG}    & Energy                    & \textbf{RY}    & Realty                    \\ \hline
\textbf{EM}    & Electrical Machinery      & \textbf{SC}    & Semiconductor             \\ \hline
\textbf{EU}    & Energy \& Utilities       & \textbf{ST}    & Steel                     \\ \hline
\textbf{FB}    & Food \& Beverages         & \textbf{SU}    & Securities                \\ \hline
\textbf{FN}    & Finance                   & \textbf{TC}    & Telecom                   \\ \hline
\textbf{GD}    & Gold                      & \textbf{TD}    & Trade                     \\ \hline
\textbf{HC}    & Health Care               & \textbf{TE}    & Transport \& Equipment    \\ \hline
\textbf{HG}    & Household Goods           & \textbf{TP}    & Transport                 \\ \hline
\textbf{HT}    & Hotel \& Tourism          & \textbf{TS}    & Trade \& Services         \\ \hline
\textbf{ID}    & Industries                & \textbf{TT}    & Travel \& Tourism         \\ \hline
\textbf{IF}    & Infrastructure            & \textbf{TX}    & Textiles                  \\ \hline
\textbf{IP}    & Industrial Production     & \textbf{UT}    & Utilities                 \\ \hline
\textbf{IS}    & Insurance                 & \textbf{WS}    & Wholesale                 \\ \hline
\end{tabular}
\end{tiny}
\label{Table:sectoral_index}
\end{table}


\subsection{Correlation coefficient and the distance metric}
If $r_{1...N}$ represents the return of $N$ sectors, which is calculated as
$r_i( \tau ) = \ln P_i( \tau )- \ln P_i( \tau -1)$, where $P_i(\tau )$ is the adjusted closure price of sector $i$ in day $\tau$, then the equal time Pearson correlation coefficients between sectors $i$ and $j$ is defined as
\begin{eqnarray}
\rho_{ij} = \frac{\langle r_i r_j \rangle - \langle r_i \rangle \langle r_j \rangle}{\sqrt{ [\langle r^{ 2}_i \rangle - \langle r_i \rangle^2][\langle r^{ 2}_j \rangle - \langle r_j \rangle^2]}}, 
\label{Eq:pearson_corr}
\end{eqnarray}
where $\langle...\rangle$ represents the expectation. We  use $\boldsymbol{\rho}$ to denote the return correlation matrix.

Following a standard procedure in the literature, we construct the distance metrix from the correlation coefficients using the following transformation,
	 $d_{ij} = \sqrt{2 (1- \rho_{ij})}$,
where $2 \geq d_{ij} \geq 0$. All elements of the matrix $d_{ij}$ are ``ultrametric" \cite{Rammal:86,Mantegna:99,OnnelaI:03}).




\subsection{Eigenvector centrality} 
To analyze the influence of a sector in the whole network, the ranking of the sectors is measured by the eigenvector centrality. It is not necessary that a sector with high eigenvector centrality is highly linked but the sector might have few but important links. Given an $N\times N$ matrix $\boldsymbol{A}$, the eigenvector centrality is defined as an $N\times 1$ vector $\boldsymbol{x}$, which solves
\begin{eqnarray}
		\boldsymbol{A x} = \lambda_m \boldsymbol{x},
\end{eqnarray}
where $\lambda_m$ is the dominant eigenvalue of $\boldsymbol{A}$.

In general, almost all pair-wise correlations are positive. However, in rare cases (e.g., Gold sector in Canada), certain sectors show mild negative correlations with other sectors.
We consider the absolute value of the correlation matrix $|\boldsymbol{\rho}|$ for computing the eigenvector centrality, since according to the Perron-Frobenius theorem, a real square matrix with positive entries has a unique largest real eigenvalue and the corresponding eigenvector has strictly positive components. Finally, we normalize the centrality vector $\boldsymbol{x}$ such that
$\sum_i x_i$ = 1.

We consider a further modification of the centrality measure to identify the core-periphery structure in a binary fashion. Instead of the level values of the correlation coefficients, we consider $\boldsymbol{\rho}^{c}$, where $c$ is a sufficiently large \textit{even} number, since this transformation would make the many weak correlations have asymptotically zero weights while maintaining positive signs.
We found that $c  = 2^5 = 32$ is the lowest value, which gives reasonably good estimates of the backbone of the minimum spanning tree. Hence, we present results for $c=32$ although, in principle, one can use higher values as well. To determine  the core sectors of a country, we then construct a threshold value $\theta_e$, as a fixed percentage of the coefficient of variation (standard deviation/mean) for  the eigenvector centralities. If the sectoral centrality is above the threshold value $\theta_e$, then the sector is considered as core, otherwise not. 



\subsection{Multidimensional scaling}
To analyze the similarity among different sectors in terms of distances ($d_{ij}$),  geometrical maps are generated using MDS for each of 27 countries, where each sector corresponds to a set of coordinates in a multi-dimensional space. The concept behind MDS is to represent two similar sectors as two sets of coordinates that are close to each other, and two sectors behaving differently are placed far apart in the space \cite{Borg:05}.
Given $d_{ij}$, the aim of MDS is to generate $N$ vectors $y_1,...,y_N \in \Re^q$, such that

\begin{eqnarray}
\Arrowvert y_i  -  y_j \Arrowvert \approx d_{ij} \hspace{0.3in}\forall i, j\in N, 
\end{eqnarray}
where $\Arrowvert . \Arrowvert$ represents vector norm. To plot the vectors $y_i$ in the form of a map, the embedding dimension $q$ is chosen as $2$. 
Generally, MDS can be obtained through an optimization problem, where ($y_1,...,y_N$)  is the solution of the problem of minimization of a cost function, such as

\begin{eqnarray}
\min_{y_1,...,y_N} \sum_{i<j} (\Arrowvert y_i - y_j \Arrowvert - d_{ij})^q.
\end{eqnarray}
 

\subsection{Minimum spanning tree (MST)} 
MST is a clustering algorithm.
By giving distance matrix $ d_{ij}$ as input, MSTs are constructed for $N$ sectors for each of the 27 countries, which are connected, undirected graphs such that all the $N$ sectors are connected together with the minimal total weighting for its $N-1$ edges, i.e., the total distance is minimum.

\subsection{Linear regression}
For relating the size with the variation of centrality, we employ the standard econometric technique of ordinary least squares.
Let us assume that the model to explain $x_i$ data-points, for $i=1, \dots , N$, is given by the following 
\begin{equation}
y_i=\beta_0+\beta_1 x_i+\epsilon_i,
\label{eq:regression}
\end{equation}
where $x$ is the explanatory variable, $y$ is the dependent variable and $\epsilon$ denotes error terms. The ordinary least squares method minimizes the sum of the squared errors, to estimate the coefficients $\beta$. Throughout our regression analyses, we have used scaled variables: [variable-mean(variable)]/ standard deviation(variable).
We carried out the estimation exercise using the MATLAB and R software packages. 

\section{Results}

In this section, we describe the three main results. Firstly, the return correlation network closely mimics the actual production network and so the core-periphery structure of the return correlation network is closely associated with the relative sizes of the sectors. Secondly, based on the findings for two periods of the empirical data (calm and volatile periods), our results are robust and hence universal, with respect to time.
 Finally, the core sectors which are typically very large in size, drives the market mode of the returns and hence, is riskier than the peripheral sectors as observed in minimum variance portfolio management.

\subsection{Financial fluctuations and economic fundamentals} 
\begin{figure}
\centering
	\includegraphics[width=0.85\linewidth]{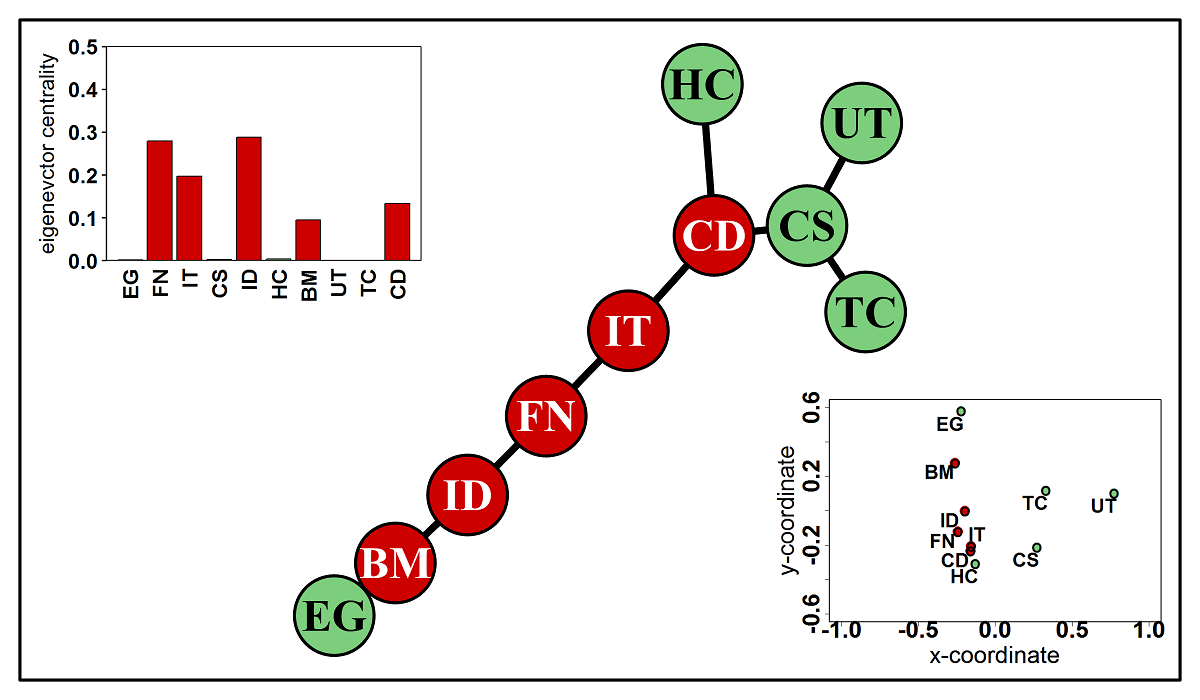}
	\includegraphics[width=0.95\linewidth]{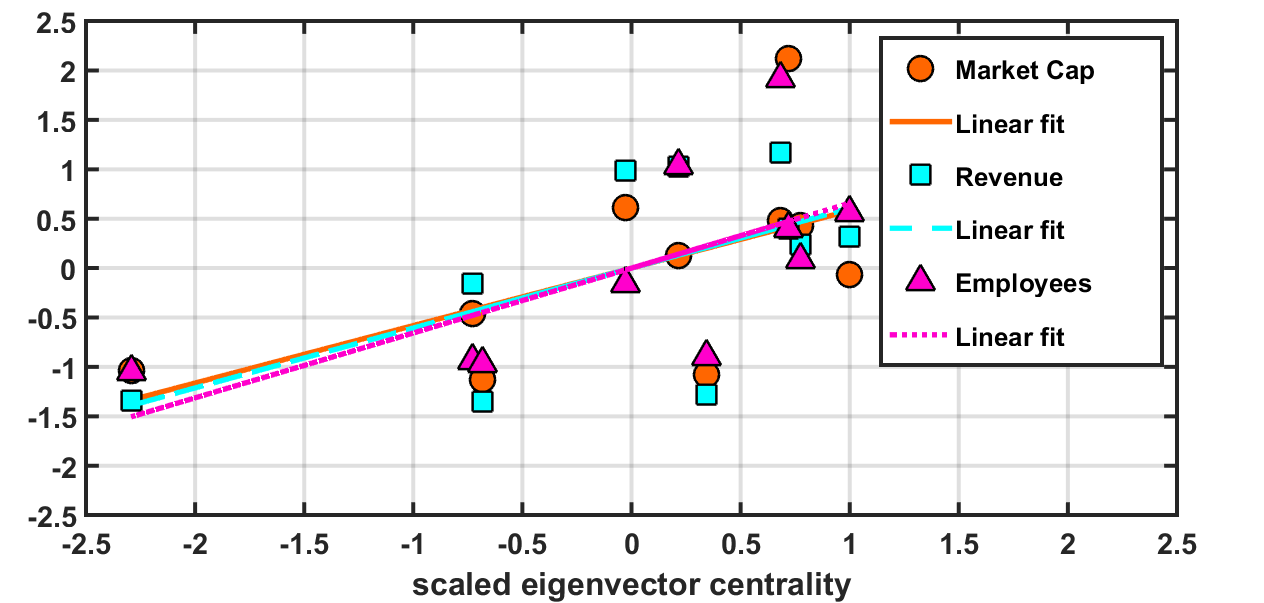}
\caption{(Color online) Results for USA: (\textit{Upper}) Identification of the sectors that are in the core (red) and periphery (pale green) of the minimum spanning tree, where the nodes represent different sectors; sectoral abbreviations given in the table \ref{Table:sectoral_index}.  \textit{Top Left Inset}: Eigenvector centralities of $\rho^{32}$. \textit{Lower Right Inset}: Multidimensional scaling, where the different sectors are plotted as coordinates in a map. (\textit{Lower}): Linear regressions of scaled eigenvector centrality with scaled market capitalization (orange filled circles), scaled revenue (cyan filled squares), and scaled number of employees (magenta filled up-triangles). The best fits  (linear regressions) are plotted as lines for market capitalization (orange solid), revenue (cyan long dashed) and employees (magenta short dashed). The variables have been scaled so that they can be plotted and compared in the same figure. \\
}
\label{fig:US_network}
\end{figure}

Given the return correlation  $\boldsymbol{\rho}$, we computed the modified eigenvector centralities to find the core sectors of the countries, and to visualize the co-movements and clusters of sectors based on return correlations, we applied two clustering algorithms, viz., MDS and MST. Fig. \ref{fig:US_network} (\textit{Upper}) shows the MST. Fig. \ref{fig:US_network} (\textit{Upper Left Inset})
shows that using the eigenvector centrality, we can identify that out of 10 sectors of the USA, 5 sectors constitute the core of the economy, viz., Finance (FN), Information Technology (IT), Industries (ID), Basic Materials (BM) and Consumer Discretionaries (CD) (see table \ref{Table:sectoral_index} in Sec. \ref{subsec:data} for names of the sectors). 
Fig. \ref{fig:US_network} (\textit{Lower Right Inset}) shows the MDS. The MST generates a core-periphery structure based on minimizing the distance between correlated sectors, and since it is a hierarchical clustering method, similar sectors can be found close to each other (or in one branch). Similarly, closer the sectors are placed on the MDS map, more correlated (similar) they are; farther they are placed on the map, less correlated they are.  

There are two major observations: First, the MST shows that all core sectors form a chain or the ``backbone'' in the tree (see Fig. \ref{fig:US_network} (\textit{Upper})). Similarly, the MDS also reiterates the same information: 
the core sectors, as identified by the modified eigenvectors centrality, belong to one cluster in the MDS (see Fig. \ref{fig:US_network} (\textit{Lower Right Inset})); all sectors with negligible centrality are spaced in the periphery -- far away from the core -- in the MDS. 
Thus, our method of the modified centrality to extract the core sectors is reinforced by the clustering algorithms, indicating the robustness of our findings. Second, the MST built from the return correlation matrix, contains information about the actual production structure of the economy. For example, Energy (EG) is most closely related to Basic Materials (BM), which in turn is related to Industries (ID), and so on. On the other end of the MST, Consumer Staples (CS) is connected to Telecom (TC) sector,  Utilities (UT) and Consumer Discretionary (CD). Again, this qualitative feature is quite robust, as observed in almost all the countries analyzed.

More importantly, we show that the core-periphery structure based on the return correlation matrix, $\boldsymbol{\rho}$, has an intriguing relationship with the relative sizes of the sectors.
In order to demonstrate and establish the relationship, we study the variations in the eigenvector centralities of the return correlation matrix, and exploit the variations in three major variables,
viz., aggregate market capitalization, aggregate revenue and the aggregate employment. We have described in Sec. \ref{subsec:data} how we constructed the sector-level data by aggregating the company-level data. In Fig. \ref{fig:US_network} (\textit{Lower}), we plot the linear regressions of scaled eigenvector centrality with the (scaled) market cap, revenue and employees for the USA. We have performed similar analyses for the other countries, and tabulated the results in the Supplementary material. Detailed analyses and tables suggest that generally, such a mapping exists for almost all countries.

\begin{figure*}
\centering
	\includegraphics[width=\linewidth]{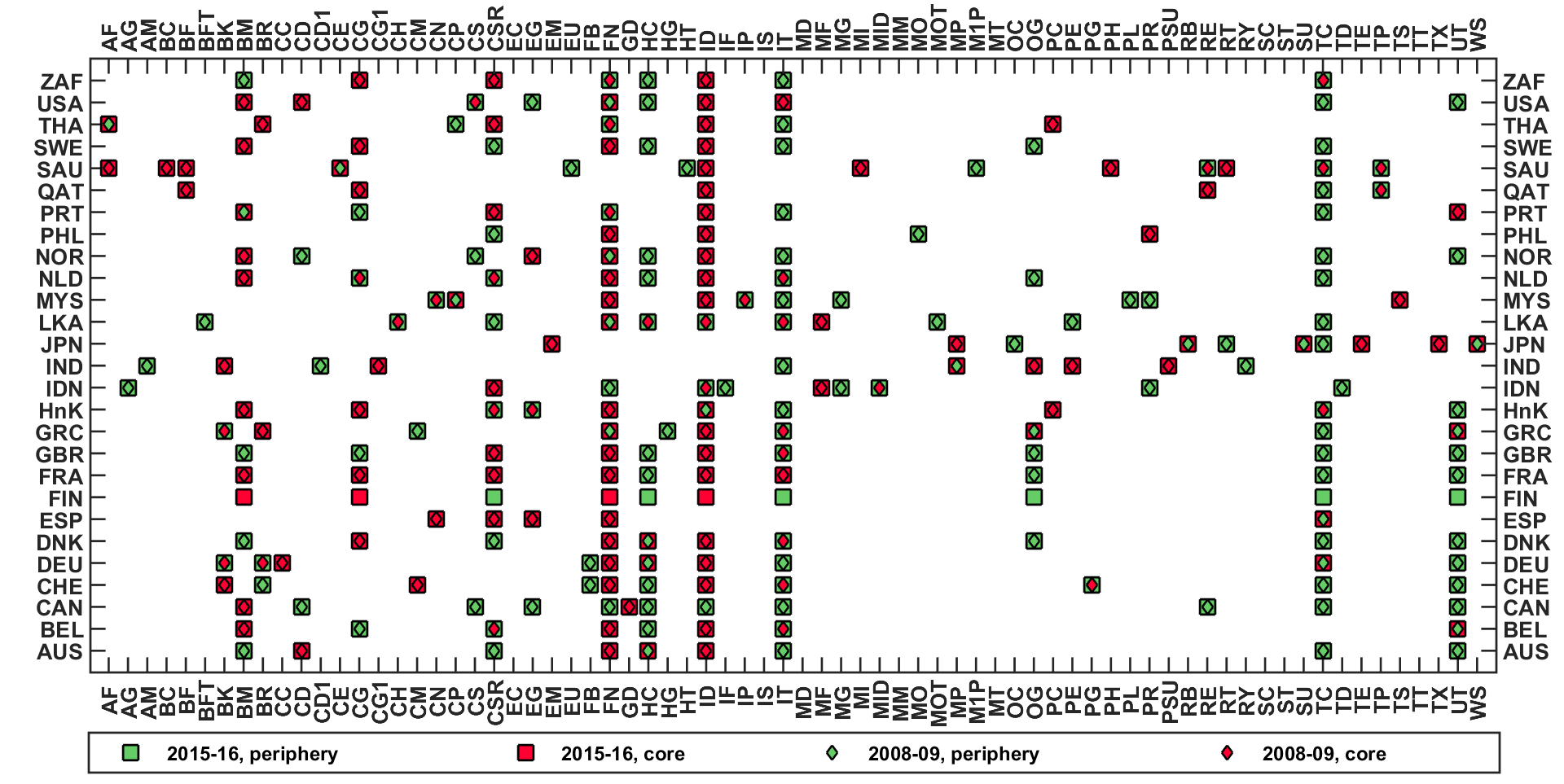}
\caption{(Color online) Sectoral dynamics and core-periphery structure: Chart of the 72 sectors (horizontal axis) in the 27 countries (vertical axis), showing the core (red) and periphery (pale-green) structures for the years 2008-09 (diamonds) and 2015-16 (squares), and their changes over time. The sector abbreviations can be found in the table \ref{Table:sectoral_index} in Sec. \ref{subsec:data}. Visual inspection reveals that sectors FN and ID are frequently occurring in the core/backbone, across almost all countries.}
\label{fig:core_periphery}
\end{figure*}

\begin{figure*}
\centering
	\includegraphics[width=\linewidth]{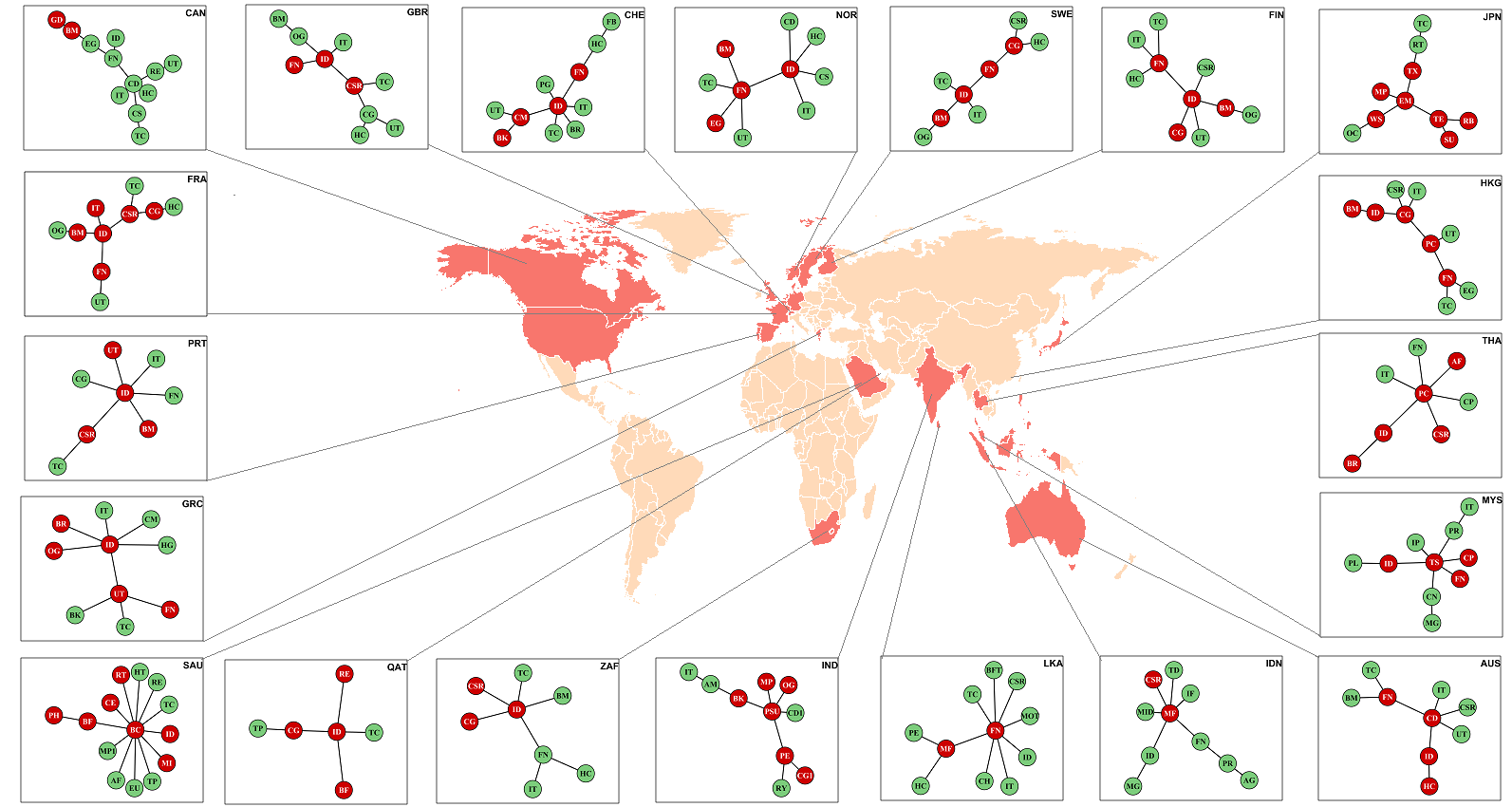}
\caption{(Color online) Minimum spanning trees for 20 countries out of the 27 countries (shown in peach) that are being studied across the globe. The core sectors are colored red (darker shade), while the sectors in the periphery are in pale green (lighter shade). The sector abbreviations can be found in the table \ref{Table:sectoral_index} in Sec. \ref{subsec:data}.}
\label{fig:world_networks}
\end{figure*}

Fig. \ref{fig:core_periphery} shows the core-periphery structure for all countries. As can be seen, there are at least two sectors in the core for all countries, but the core-periphery structure often changes with time (when compared for the periods 2008-09 and 2015-16). Thus, the relative importance of the sectors does change with time, and the sectoral dynamics and co-movements may convey deeper insight about the aggregate macro-level dynamics.
In Fig. \ref{fig:world_networks}, we present similar MSTs (with the core/backbone colored in red) for 20 other countries, elucidating the core-periphery structures. 

\begin{figure}
\centering
	\includegraphics[width=0.85\linewidth]{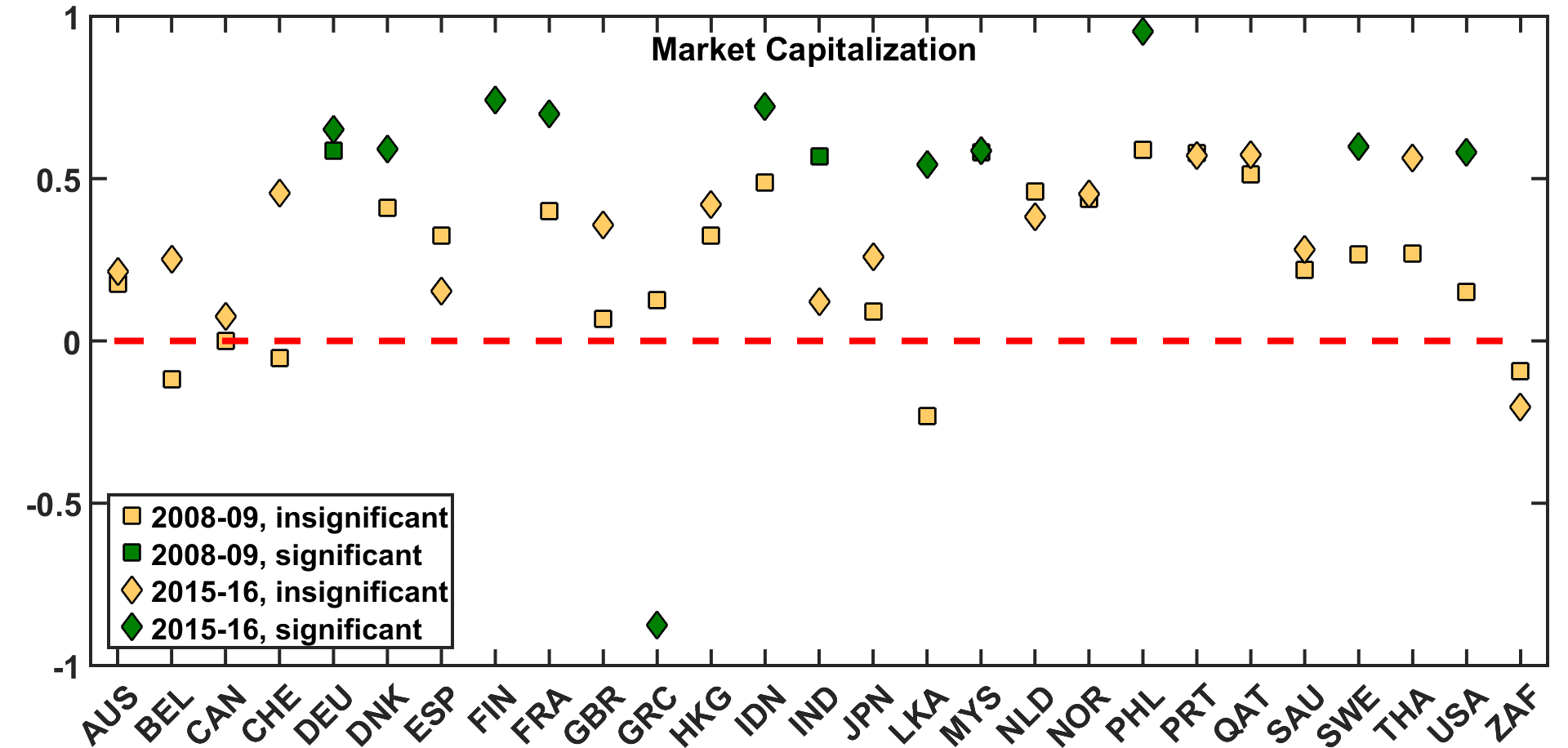}\\
	\includegraphics[width=0.85\linewidth]{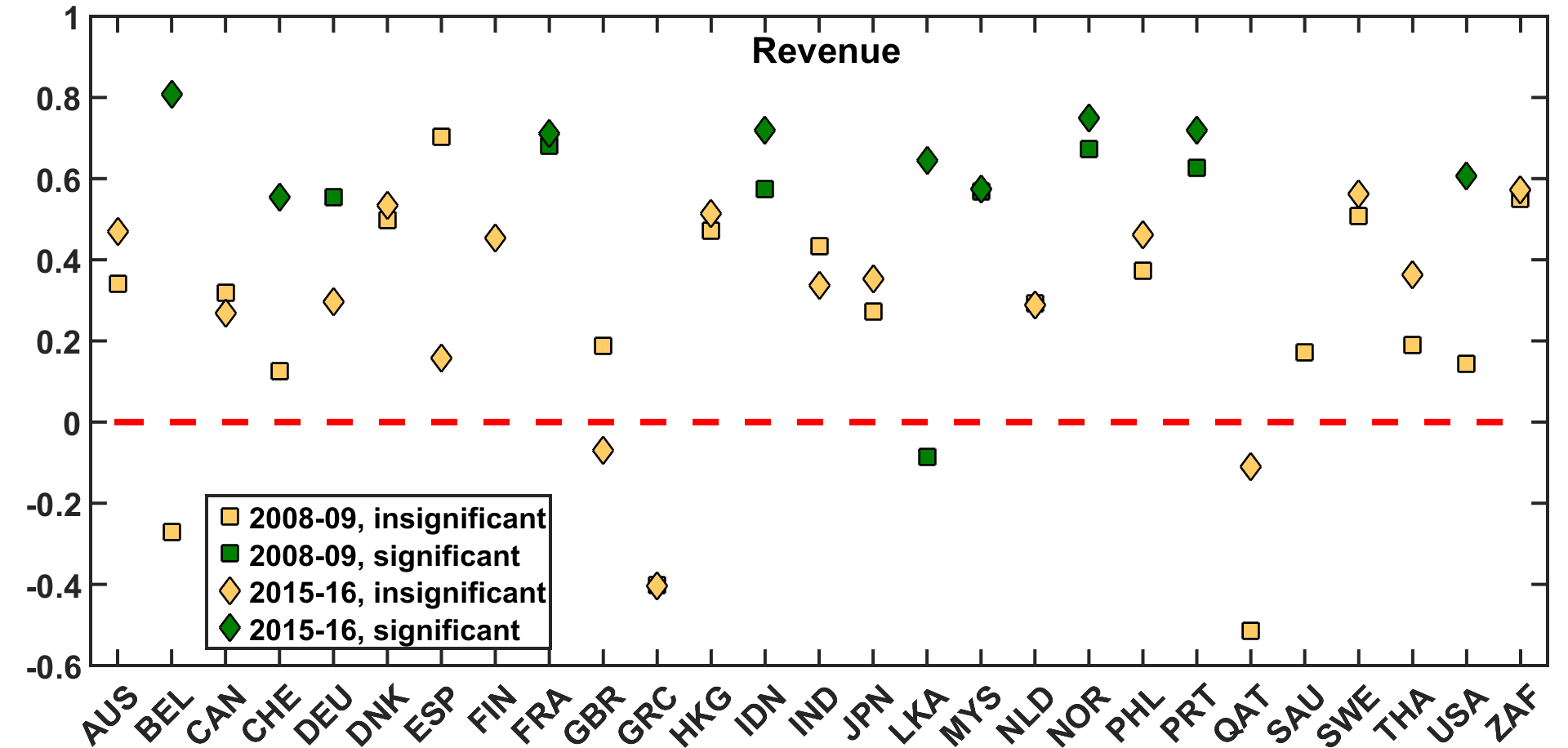}\\
	\includegraphics[width=0.85\linewidth]{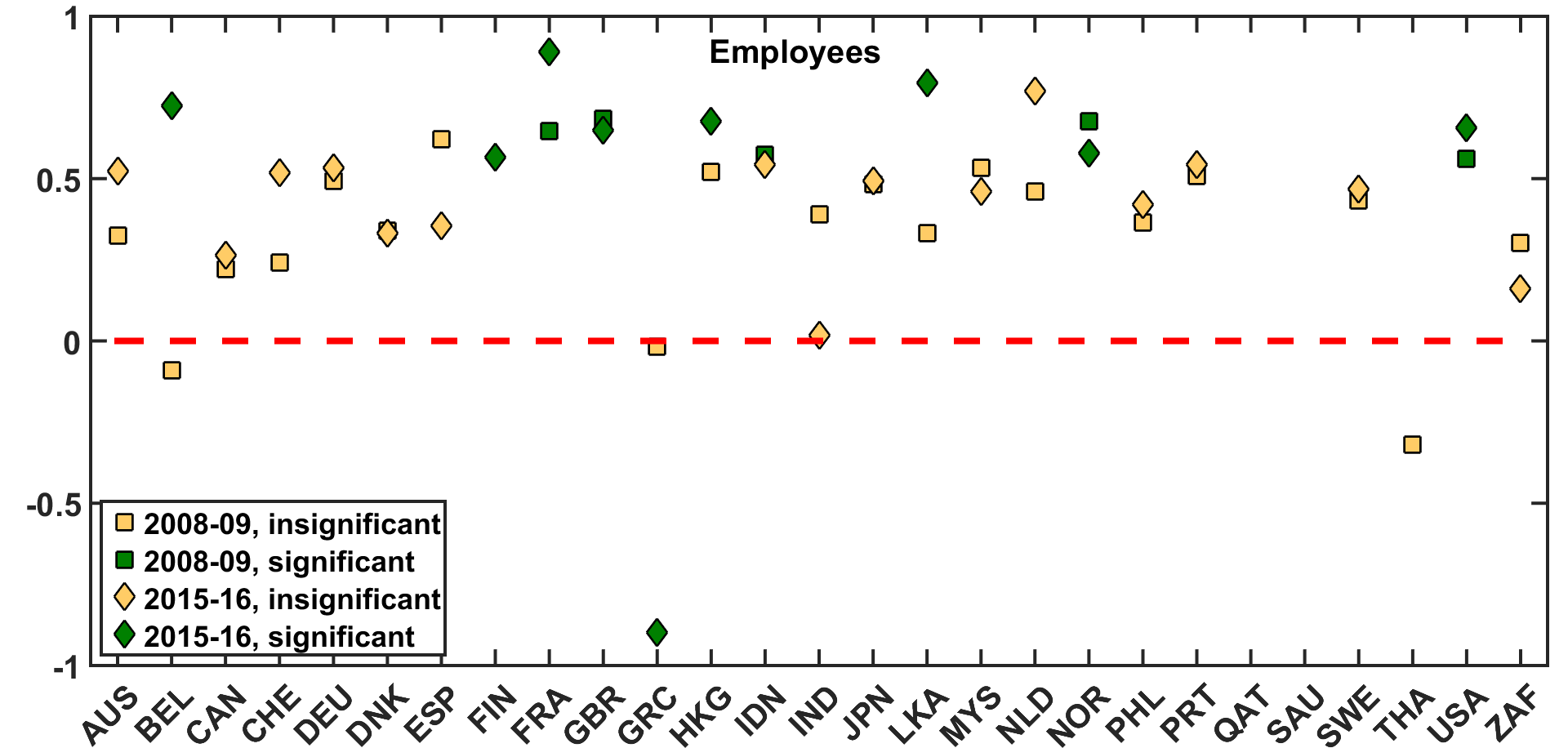}\\
\caption{(Color online) Comparison of the regression results (estimates of $\beta_1$ using Eq. \ref{eq:regression}) to explain variation in the sectoral eigenvector centralities ($y$) by the variation in sector-level macro data ($x$). \textit{Upper}: market capitalization, \textit{Middle}: Revenue, \textit{Lower}: Employees, for the years 2008-09 and 2015-16. Detailed estimation results are given in the tables in Supplementary material.}
\label{fig:regression}
\end{figure}
Fig. \ref{fig:regression}
shows the results of regressing the sectoral eigenvector centralities on the sector-level aggregate market capitalization, revenue, and employees, for the years 2008-09 and 2015-16.
As we see in Fig. \ref{fig:regression} (\textit{Upper}), for 2015-16, the coefficient $\beta_1$ for market capitalization 25 out of 27 countries are positive, and 11 out of those 27 countries have statistically significant relationships.
The two countries which have very mildly negative relationships, are Greece (significant) and South Africa (insignificant).
For 2008-09, the coefficient $\beta_1$ for 22 out of 26 countries are positive, and 3 out of those 26 countries have statistically significant relationships. Also, the countries Belgium, Switzerland, South Africa and Sri Lanka have negative relationships. In Fig. \ref{fig:regression} (\textit{Middle}), for 2015-16, the coefficient $\beta_1$ for revenue, 23 out of 26 countries are positive, and 9 out of those 26 countries have statistically significant relationships.
The three countries which have negative (and  statistically insignificant) relationships, are Greece,  Qatar and United Kingdom.
For 2008-09, the coefficient $\beta_1$ for 22 out of 26 countries are positive, and 9 out of those 24 countries have statistically significant relationships. Finally, in Fig. \ref{fig:regression} (\textit{Lower}), for 2015-16, the coefficient $\beta_1$ for employees, 23 out of 24 countries are positive, and 9 out of those 24 countries have statistically significant relationships.
The only country which has negative (and  statistically significant) relationship, is Greece.
For 2008-09, the coefficient $\beta_1$ for 21 out of 24 countries are positive, and 5 out of those 24 countries have statistically significant relationships. For detailed statistical values of regressions performed on the sector-level  aggregate data, please refer to the text in Supplementary material.

There is  already an existing finding that centralities in input-output networks are closely related to the relative sizes of the corresponding nodes (see Ref. \cite{Acemoglu_12}). 
However, here we further show that the centralities based on nominal return fluctuations are related to relative size, i.e., the return network is also very closely related to the underlying size effects. An immediate corollary is that the core sectors of the return correlation network are also economically big, and hence, the market
effect of the correlations are driven by the sectors, which have very high market capitalization (or other indicators like revenue and employment).




\subsection{Robustness: volatile period versus calm period}

\begin{figure}
\centering
	
	\includegraphics[width=0.95\linewidth]{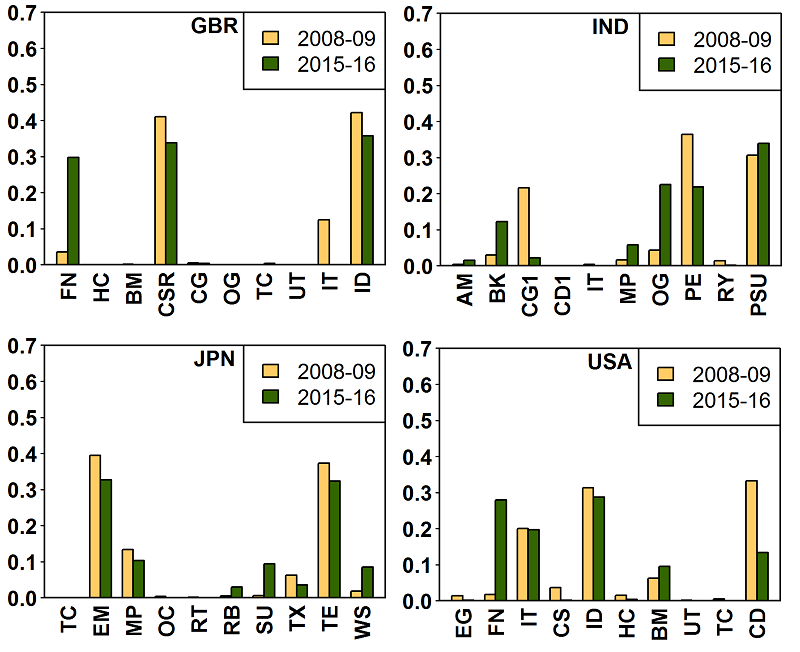}
\caption{(Color online) Sectoral dynamics and robustness: The comparison of the eigenvector centralities for the years 2008-09 (light orange) and 2015-16 (dark green) for four countries. \textit{Upper Left}: United Kingdom (GBR), \textit{Upper Right}: India (IND), \textit{Lower Left}: Japan (JPN), \textit{Lower Right}: United States of America (USA). }
\label{fig:robustness}
\end{figure}

We studied the dynamics of a total of 72 sectors across 27 countries, covering both developed and developing economies. Using methods of modified eigenvector centrality, MDS and MST, we can find the core-periphery structure of all the economies. Fig. \ref{fig:core_periphery} showed the core-periphery structure of all the countries, and indicated that most of the sectors do not change much in the core-periphery structure during the periods of market turbulence, as well as relative calmness. There are of course, some sectors who were core in a volatile period, became the peripheral ones in the calm period, and vice versa. 
Fig. \ref{fig:robustness} shows the comparison among the modified eigenvector centralities for the years 2008-09 and 2015-16, for the four countries: United Kingdom, India, Japan, and United States of America, as examples.
The relative importance of each sector can be compared for the volatile and calm period. Certainly the sectoral dynamics are interesting to note in the different countries, and may help in taking important policy decisions in economic growth and development.



\begin{figure}
\centering
   \includegraphics[width=0.95\linewidth]{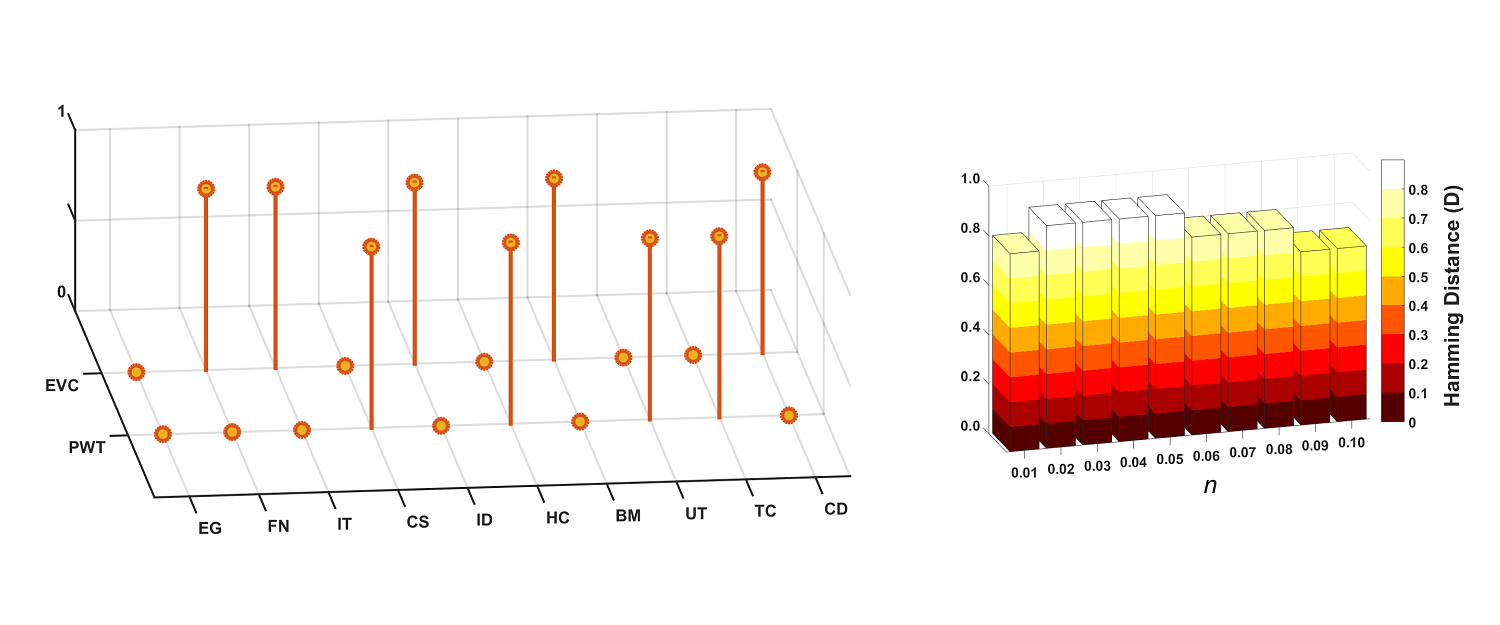}\\
	\includegraphics[width=0.85\linewidth]{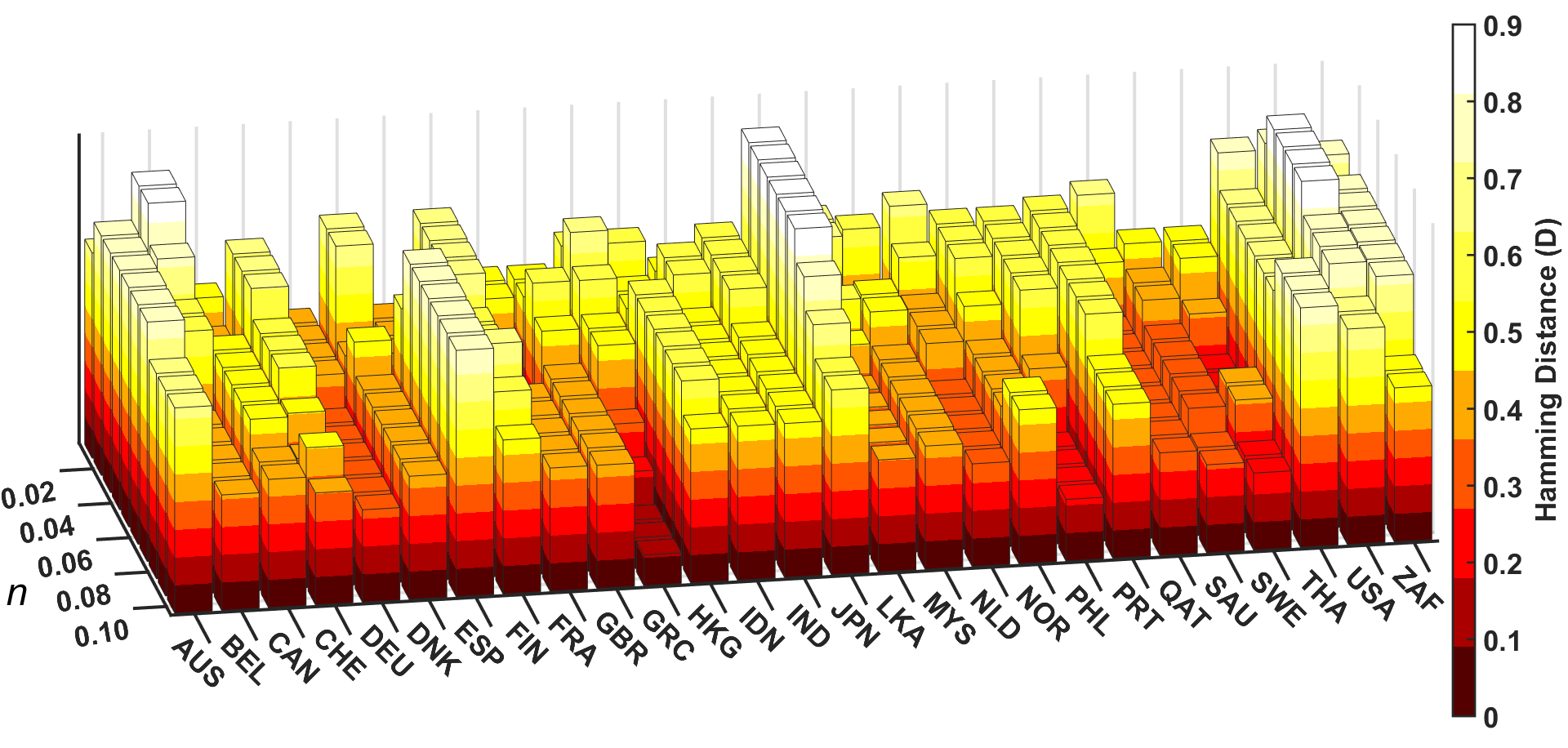}
\caption{(Color online) {\it Upper Left}: Relationship between the bit-strings of sectoral centralities (EVC) and their corresponding inclusion in the portfolio (PWT) for the different sectors of the USA. The threshold values $\theta_e$ and  $\theta_p$ (as $2\%$ of the coefficient of variations of EVC and PWT, respectively) would determine, whether the sector is central or not (EVC is $0$ or $1$), or whether the corresponding sector would appear in the optimal portfolio or not (PWT is $0$ or $1$). The labels for the different sectors are given in table \ref{Table:sectoral_index}. {\it Upper Right}: The Hamming distance $D$ computed from the bit-strings EVC and PWT, against the different values of $n$ (percentage) of the coefficient of variations of EVC and PWT, respectively, which determine the threshold values $\theta_e$ and  $\theta_p$. {\it Lower}: The Hamming distance $D$ computed from the bit-strings EVC and PWT, against the different values of $n$, for the different countries, plotted as a 3D-bar.}
\label{fig:optimal}
\end{figure}

\subsection{Constructing the minimum risk portfolio} In this part, we study how the sectoral centralities influence the aggregate risk of a portfolio. For the purpose of simple exposition, we compute the
benchmark model of Markowitz portfolio selection with the sectoral return data. Assuming rational investors with risk-aversion, the investors will minimize
\begin{equation}
w'\Sigma w-\Theta R'w,
\end{equation}
with respect to the weight vector $w$, where $\Sigma$ is the covariance matrix of the sectoral returns, $R'$ is the expected return vector and $\Theta$ is a
parameter which denotes the risk appetite of the investor. We set a short-selling constraint ($w_i \ge$ 0) and $\Theta$ equals to zero for finding the minimal risk portfolio which will entail a convex combination of sectoral returns (the other extreme would lead to a corner solution). 

Our main observation in this part is that the optimal weight vector, $w^*$, is negatively related to the eigenvector centralities, i.e., if a sector is very ``central'' in the return correlation network $\boldsymbol{\rho}$, then it is less likely to appear in the optimal portfolio with the minimum risk (and no short selling). We demonstrate this in a naive way: we construct threshold values $\theta_e$ and  $\theta_p$, as a fixed percentage (say $n\%$ of the coefficient of variation (standard deviation/mean) for both the eigenvector centralities, as well as the minimum risk portfolio weights, respectively. These threshold values $\theta_e$ and  $\theta_p$ would determine, respectively, whether the sector is central or not (i.e., $0$ or $1$), or whether the corresponding sector would appear in your optimal portfolio or not (i.e., $0$ or $1$). So, for the vector of sectors, we would have two strings of $0$'s or $1$'s corresponding to the centrality vector (EVC) and the optimal weight vector (PWT), respectively. The Hamming distance $D$ between any two bit-strings of equal length, is the number of positions at which the corresponding bits are different. So, the Hamming distance between the two strings EVC and PWT would tell how significant the observation is for a particular country; higher the value of $D$, better the conformity. The sector which is central (i.e., $1$) would not appear in your portfolio (i.e., $0$), and so for any country the ideal finding would be that $D$ is unity.
The choice of the threshold(s), $\theta_e$ and $\theta_p$, equaled by the percentage(s) ($n$) of the coefficient of variation(s) in the vectors EVC and PWT,  would be important for determining the Hamming distance $D$ between the strings for any country (see Fig. \ref{fig:optimal} (\textit{Upper})) for the USA. We can optimize the value of $D$ against the percentage $n$, for all the countries, as shown in Fig. \ref{fig:optimal} (\textit{Lower}). We found that $n=2$, i.e. 2\% was an optimal threshold value $\theta_e$ for most countries, which we then used to distinguish between the core and periphery sectors.
Combined with the finding that core sectors in the return correlation network are bigger in size, the above finding implies that peripheral sectors contribute to lower risk
of a diversified portfolio.

\section{Summary and conclusion}


In this paper, we have analyzed financial and economic data for 27 countries at the sector level.  
We show that the variation in the centrality in the return correlation matrix across sectoral indices, can be 
explained by the size dispersion across the sectors.
This finding indicates that financial fluctuations are mapped to the macroeconomic fundamentals. From the perspective of portfolio optimization, we show that
the very big sectors that are also highly central in the return network, rarely appear in a risk-minimizing portfolio. Essentially, such sectors constitute the
main drivers of the market-wide fluctuations. In summary, our study sheds light on: (a) the mapping between the joint evolution of the financial variables and the underlying macroeconomic fundamentals, and (b) extracting information about the individual influences on aggregate risk from sector-level, disaggregated time-series data.

Methodologically, we provide a way to extract the core-periphery structure of the correlation networks in a binary fashion.
As a result, the generic rule of thumb we come up with is that size is an important causal factor even behind financial fluctuations. 
We attribute significant importance to this finding as it provides a way to exactly pin down the sectors, which are main drivers of financial fluctuations through the size effect.
The way return series are constructed, the size differential of the prices across the sectoral indices, should disappear due to the normalization. The fact that
the co-movements are still tied to the fundamentals is therefore intriguing. As our results suggest, the finding is considerably robust across countries.
An illuminating exception is Greece showing an exact opposite relationship, which has been known to possess weak economic fundamentals along with severe crises in the financial markets in the recent times. In both periods, economically large (either in terms of market capitalization or revenue or employment) sectors in Greece are at the periphery of the return correlation networks, which constitute an inverted relationship between the economy and the financial networks. 

We have also shown that the relative importance of the sectors may change significantly over time although some sectors like finance and industry are at the core of a significant fraction of countries. In general, our results indicate that the core may not be very stable. Possible reasons could be sectoral competition in terms of productivity and innovation and the resultant evolution \cite{Acemoglu_16}. The emergence of the core-periphery structure changes the complexity of the financial markets and has implications of the 
pricing of risk in the economy \cite{Battiston_16}. Our work indicates the potentials of using a binary characterization to reduce the computational burden by introducing proper identification of the country-specific core sectors, as opposed to considering the full network.

To conclude, we note that the recent applications of network theory in the macroeconomics literature has focused mostly on 
studying the dynamics of real economic quantities \cite{JEDC_15}, whereas the 
relevant finance literature has focused on the dynamics of nominal quantities \cite{Mantegna_Stanley_book}. The present work may provide a linkage between the two.
In other words, we make the point that the oft-quoted quips `too-big-to-fail' and `too-interconnected-to-fail' may not be as different as is currently thought of \cite{Acemoglu_15}.

\section*{Acknowledgment}
ASC acknowledges the support by the institute grant (R\&P), IIM Ahmedabad. BG acknowledges FPM fellowship provided by IIM Ahmedabad. AC and KS acknowledge the support by grant number BT/BI/03/004/2003(C) of Govt. of India, Ministry of Science and Technology, Department of Biotechnology, Bioinformatics division, and University of Potential Excellence-II grant (Project ID-47) of the Jawaharlal Nehru University, New Delhi.
KS acknowledges the University Grants Commission (Ministry of Human Research Development, Govt. of India) for her senior research fellowship.

\bibliographystyle{IEEEtr}
\bibliography{main}

\newpage

\newpage
\section*{Supplementary material}
All detailed regression tables are provided below.
\begin{table}[H]
\caption{Regression table: Dependent variable is the scaled centrality and the independent variable is the scaled market capitalization (2015-16). *** : significant at 1\%, **: at 5\%, *: at 10\%.}
\begin{tabular}{|l|c|c|c|c|c|c|c|}
\hline
\textbf{Countries} & \textbf{$\beta_0$}                                               & \textbf{Tstat} & \textbf{Pvalue} & \textbf{$\beta_1$}                                               & \textbf{Tstat} & \textbf{Pvalue} & \textbf{Rsquare} \\ \hline
Australia    & \begin{tabular}[c]{@{}l@{}}0.0000\\   (0.3481)\end{tabular} & 0.0000                                             & 0.9999 & \begin{tabular}[c]{@{}l@{}}0.2131\\   (0.3692)\end{tabular}  & 0.5773  & 0.5817    & 0.0454  \\ \hline
Belgium      & \begin{tabular}[c]{@{}l@{}}0.0000\\   (0.3695)\end{tabular} & 0.0000                                             & 0.9999 & \begin{tabular}[c]{@{}l@{}}0.2517\\   (0.3951)\end{tabular}  & 0.6371  & 0.5475    & 0.0633  \\ \hline
Canada       & \begin{tabular}[c]{@{}l@{}}0.0000\\   (0.3018)\end{tabular} & 0.0000                                             & 0.9999 & \begin{tabular}[c]{@{}l@{}}0.0758\\   (0.3153)\end{tabular}  & 0.2404  & 0.8148    & 0.0057  \\ \hline
Denmark      & \begin{tabular}[c]{@{}l@{}}0.0000\\   (0.2703)\end{tabular} & 0.0000                                             & 0.9999 & \begin{tabular}[c]{@{}l@{}}0.5918\\   (0.2849)\end{tabular}  & 2.0767  & 0.0714*   & 0.3502  \\ \hline
Finland      & \begin{tabular}[c]{@{}l@{}}0.0000\\   (0.2246)\end{tabular} & 0.0000                                             & 0.9999 & \begin{tabular}[c]{@{}l@{}}0.7425\\   (0.2368)\end{tabular}  & 3.1353  & 0.0139**  & 0.5513  \\ \hline
France       & \begin{tabular}[c]{@{}l@{}}0.0000\\   (0.2523)\end{tabular} & 0.0000                                             & 0.9999 & \begin{tabular}[c]{@{}l@{}}0.7003\\   (0.2394)\end{tabular}  & 2.7751  & 0.0241**  & 0.4904  \\ \hline
Germany      & \begin{tabular}[c]{@{}l@{}}0.0000\\   (0.2544)\end{tabular} & 0.0000                                             & 0.9999 & \begin{tabular}[c]{@{}l@{}}0.6516\\   (0.2681)\end{tabular}  & 2.4299  & 0.0412**  & 0.4246  \\ \hline
Greece       & \begin{tabular}[c]{@{}l@{}}0.0000\\   (0.1617)\end{tabular} & 0.0000                                             & 0.9999 & \begin{tabular}[c]{@{}l@{}}-0.8759\\   (0.1705)\end{tabular} & -5.1368 & 0.0008*** & 0.7673  \\ \hline
Hong Kong    & \begin{tabular}[c]{@{}l@{}}0.0000\\   (0.3042)\end{tabular} & 0.0000                                             & 0.9999 & \begin{tabular}[c]{@{}l@{}}0.4208\\   (0.3207)\end{tabular}  & 1.3123  & 0.2257    & 0.1771  \\ \hline
India        & \begin{tabular}[c]{@{}l@{}}0.0000\\   (0.3329)\end{tabular} & 0.0000                                             & 0.9999 & \begin{tabular}[c]{@{}l@{}}0.1219\\   (0.3509)\end{tabular}  & 0.3474  & 0.7373    & 0.0149  \\ \hline
Indonesia    & \begin{tabular}[c]{@{}l@{}}0.0000\\   (0.232)\end{tabular}  & 0.0000                                             & 0.9999 & \begin{tabular}[c]{@{}l@{}}0.7221\\   (0.2445)\end{tabular}  & 2.9524  & 0.0183**  & 0.5214  \\ \hline
Japan        & \begin{tabular}[c]{@{}l@{}}0.0000\\   (0.3239)\end{tabular} & 0.0000                                             & 0.9999 & \begin{tabular}[c]{@{}l@{}}0.2588\\   (0.3415)\end{tabular}  & 0.7579  & 0.4702    & 0.0670  \\ \hline
Malaysia     & \begin{tabular}[c]{@{}l@{}}0.0000\\   (0.2719)\end{tabular} & 0.0000                                             & 0.9999 & \begin{tabular}[c]{@{}l@{}}0.5853\\   (0.2866)\end{tabular}  & 2.0421  & 0.0754*   & 0.3426  \\ \hline
Netherlands  & \begin{tabular}[c]{@{}l@{}}0.0000\\   (0.3292)\end{tabular} & 0.0000                                             & 0.9999 & \begin{tabular}[c]{@{}l@{}}0.3821\\   (0.3492)\end{tabular}  & 1.0941  & 0.3100    & 0.1460  \\ \hline
Norway       & \begin{tabular}[c]{@{}l@{}}0.0000\\   (0.2992)\end{tabular} & 0.0000                                             & 0.9999 & \begin{tabular}[c]{@{}l@{}}0.4516\\   (0.3154)\end{tabular}  & 1.4316  & 0.1901    & 0.2039  \\ \hline
Philippines  & \begin{tabular}[c]{@{}l@{}}0.0000\\   (0.1552)\end{tabular} & 0.0000                                             & 0.9999 & \begin{tabular}[c]{@{}l@{}}0.9537\\   (0.1736)\end{tabular}  & 5.4929  & 0.0118**  & 0.9095  \\ \hline
Portugal     & \begin{tabular}[c]{@{}l@{}}0.0000\\   (0.3133)\end{tabular} & 0.0000                                             & 0.9999 & \begin{tabular}[c]{@{}l@{}}0.5715\\   (0.3349)\end{tabular}  & 1.7061  & 0.1388    & 0.3266  \\ \hline
\end{tabular}
\label{Table:regression_markcap}
\end{table}

\begin{table}[ht]
\centering
\begin{tabular}{|l|c|c|c|c|c|c|c|}
\hline
Qatar        & \begin{tabular}[c]{@{}l@{}}0.0000\\   (0.3742)\end{tabular} & 0.0000                                             & 0.9999 & \begin{tabular}[c]{@{}l@{}}0.5723\\   (0.41)\end{tabular}    & 1.3959  & 0.2352    & 0.3275  \\ \hline
Saudi Arabia & \begin{tabular}[c]{@{}l@{}}0.0000\\   (0.2669)\end{tabular} & 0.0000                                             & 0.9999 & \begin{tabular}[c]{@{}l@{}}0.2807\\   (0.277)\end{tabular}   & 1.0133  & 0.3309    & 0.0788  \\ \hline
South Africa & \begin{tabular}[c]{@{}l@{}}0.0000\\   (0.3739)\end{tabular} & 0.0000                                             & 0.9999 & \begin{tabular}[c]{@{}l@{}}-0.2033\\   (0.3997)\end{tabular} & -0.5086 & 0.6291    & 0.0413  \\ \hline
Spain        & \begin{tabular}[c]{@{}l@{}}0.0000\\   (0.2684)\end{tabular} & 0.0000                                             & 0.9999 & \begin{tabular}[c]{@{}l@{}}0.1531\\   (0.5705)\end{tabular}  & 0.8057  & 0.1577    & 0.0234  \\ \hline
Sri Lanka    & \begin{tabular}[c]{@{}l@{}}0.0000\\   (0.267)\end{tabular}  & 0.0000                                             & 0.9999 & \begin{tabular}[c]{@{}l@{}}0.5423\\   (0.28)\end{tabular}    & 1.9363  & 0.0848*   & 0.2940  \\ \hline
Sweden       & \begin{tabular}[c]{@{}l@{}}0.0000\\   (0.2852)\end{tabular} & 0.0000                                             & 0.9999 & \begin{tabular}[c]{@{}l@{}}0.5995\\   (0.3025)\end{tabular}  & 1.9818  & 0.0879*   & 0.3594  \\ \hline
Switzerland  & \begin{tabular}[c]{@{}l@{}}0.0000\\   (0.2831)\end{tabular} & 0.0000                                             & 0.9999 & \begin{tabular}[c]{@{}l@{}}0.4543\\   (0.2969)\end{tabular}  & 1.5302  & 0.1603    & 0.2064  \\ \hline
Thailand     & \begin{tabular}[c]{@{}l@{}}0.0000\\   (0.3152)\end{tabular} & 0.0000                                             & 0.9999 & \begin{tabular}[c]{@{}l@{}}0.5643\\   (0.337)\end{tabular}   & 1.6743  & 0.1450    & 0.3184  \\ \hline
UK           & \begin{tabular}[c]{@{}l@{}}0.0000\\   (0.3132)\end{tabular} & 0.0000                                             & 0.9999 & \begin{tabular}[c]{@{}l@{}}0.3575\\   (0.3301)\end{tabular}  & 1.0830  & 0.3103    & 0.1278  \\ \hline
USA          & \begin{tabular}[c]{@{}l@{}}0.0000\\   (0.2728)\end{tabular} & 0.0000                                             & 0.9999 & \begin{tabular}[c]{@{}l@{}}0.5817\\   (0.2875)\end{tabular}  & 2.0228  & 0.0777*   & 0.3384 \\ \hline
\end{tabular}
\end{table}


\begin{table}[ht]
\centering
\caption{Regression table: Dependent variable is the scaled centrality and the independent variable is the scaled revenue (2015-16). *** : significant at 1\%, **: at 5\%, *: at 10\%.}
\begin{tabular}{|l|c|c|c|c|c|c|c|}
\hline
\textbf{Countries} & \textbf{$\beta_0$}                                               & \textbf{Tstat} & \textbf{Pvalue} & \textbf{$\beta_1$}                                               & \textbf{Tstat} & \textbf{Pvalue} & \textbf{Rsquare} \\ \hline
Australia    & \begin{tabular}[c]{@{}l@{}}0.0000\\   (0.3144)\end{tabular} & 0.0000                                             & 0.9999 & \begin{tabular}[c]{@{}l@{}}0.4704\\   (0.3335)\end{tabular}  & 1.4106  & 0.2012   & 0.2213  \\ \hline
Belgium      & \begin{tabular}[c]{@{}l@{}}0.0000\\   (0.2247)\end{tabular} & 0.0000                                             & 0.9999 & \begin{tabular}[c]{@{}l@{}}0.8084\\   (0.2402)\end{tabular}  & 3.3644  & 0.0151** & 0.6535  \\ \hline
Canada       & \begin{tabular}[c]{@{}l@{}}0.0000\\   (0.2916)\end{tabular} & 0.0000                                             & 0.9999 & \begin{tabular}[c]{@{}l@{}}0.2689\\   (0.3045)\end{tabular}  & 0.8830  & 0.3979   & 0.0723  \\ \hline
Denmark      & \begin{tabular}[c]{@{}l@{}}0.0000\\   (0.2837)\end{tabular} & 0.0000                                             & 0.9999 & \begin{tabular}[c]{@{}l@{}}0.5333\\   (0.299)\end{tabular}   & 1.7834  & 0.1123   & 0.2844  \\ \hline
Finland      & \begin{tabular}[c]{@{}l@{}}0.0000\\   (0.2988)\end{tabular} & 0.0000                                             & 0.9999 & \begin{tabular}[c]{@{}l@{}}0.4542\\   (0.3149)\end{tabular}  & 1.4422  & 0.1872   & 0.2063  \\ \hline
France       & \begin{tabular}[c]{@{}l@{}}0.0000\\   (0.236)\end{tabular}  & 0.0000                                             & 0.9999 & \begin{tabular}[c]{@{}l@{}}0.7104\\   (0.2487)\end{tabular}  & 2.8557  & 0.0212** & 0.5048  \\ \hline
Germany      & \begin{tabular}[c]{@{}l@{}}0.0000\\   (0.3203)\end{tabular} & 0.0000                                             & 0.9999 & \begin{tabular}[c]{@{}l@{}}0.2964\\   (0.3376)\end{tabular}  & 0.8778  & 0.4056   & 0.0878  \\ \hline
Greece       & \begin{tabular}[c]{@{}l@{}}0.0000\\   (0.3069)\end{tabular} & 0.0000                                             & 0.9999 & \begin{tabular}[c]{@{}l@{}}-0.4029\\   (0.3235)\end{tabular} & -1.2453 & 0.2482   & 0.1623  \\ \hline
Hong Kong    & \begin{tabular}[c]{@{}l@{}}0.0000\\   (0.2877)\end{tabular} & 0.0000                                             & 0.9999 & \begin{tabular}[c]{@{}l@{}}0.5140\\   (0.3032)\end{tabular}  & 1.6948  & 0.1285   & 0.2642  \\ \hline
India        & \begin{tabular}[c]{@{}l@{}}0.0000\\   (0.3158)\end{tabular} & 0.0000                                             & 0.9999 & \begin{tabular}[c]{@{}l@{}}0.3368\\   (0.3329)\end{tabular}  & 1.0117  & 0.3413   & 0.1134  \\ \hline
Indonesia    & \begin{tabular}[c]{@{}l@{}}0.0000\\   (0.233)\end{tabular}  & 0.0000                                             & 0.9999 & \begin{tabular}[c]{@{}l@{}}0.7193\\   (0.2456)\end{tabular}  & 2.9286  & 0.0190** & 0.5173  \\ \hline
Japan        & \begin{tabular}[c]{@{}l@{}}0.0000\\   (0.3137)\end{tabular} & 0.0000                                             & 0.9999 & \begin{tabular}[c]{@{}l@{}}0.3531\\   (0.3307)\end{tabular}  & 1.0678  & 0.3167   & 0.1247  \\ \hline
Malaysia     & \begin{tabular}[c]{@{}l@{}}0.0000\\   (0.2746)\end{tabular} & 0.0000                                             & 0.9999 & \begin{tabular}[c]{@{}l@{}}0.5739\\   (0.2895)\end{tabular}  & 1.9824  & 0.0827*  & 0.3294  \\ \hline
Netherlands  & \begin{tabular}[c]{@{}l@{}}0.0000\\   (0.3411)\end{tabular} & 0.0000                                             & 0.9999 & \begin{tabular}[c]{@{}l@{}}0.2886\\   (0.3618)\end{tabular}  & 0.7976  & 0.4512   & 0.0833  \\ \hline
Norway       & \begin{tabular}[c]{@{}l@{}}0.0000\\   (0.222)\end{tabular}  & 0.0000                                             & 0.9999 & \begin{tabular}[c]{@{}l@{}}0.7495\\   (0.234)\end{tabular}   & 3.2031  & 0.0125** & 0.5618  \\ \hline
Philippines  & \begin{tabular}[c]{@{}l@{}}0.0000\\   (0.4578)\end{tabular} & 0.0000                                             & 0.9999 & \begin{tabular}[c]{@{}l@{}}0.4625\\   (0.5118)\end{tabular}  & 0.9035  & 0.4328   & 0.2139  \\ \hline
Portugal     & \begin{tabular}[c]{@{}l@{}}0.0000\\   (0.2655)\end{tabular} & 0.0000                                             & 0.9999 & \begin{tabular}[c]{@{}l@{}}0.7187\\   (0.2838)\end{tabular}  & 2.5320  & 0.0445** & 0.5165  \\ \hline
\end{tabular}
\label{Table:regression_revenue}
\end{table}

\begin{table}[ht]
\centering
\begin{tabular}{|l|c|c|c|c|c|c|c|}
\hline
Qatar        & \begin{tabular}[c]{@{}l@{}}0.0000\\   (0.4536)\end{tabular} & 0.0000                                             & 0.9999 & \begin{tabular}[c]{@{}l@{}}-0.1102\\   (0.4969)\end{tabular} & -0.2219 & 0.8352   & 0.0121  \\ \hline
Saudi Arabia &               \begin{tabular}[c]{@{}l@{}} - \\ -\end{tabular}                                               &                                                    &        &                                                              &         &          &         \\ \hline
South Africa & \begin{tabular}[c]{@{}l@{}}0.0000\\   (0.3129)\end{tabular} & 0.0000                                             & 0.9999 & \begin{tabular}[c]{@{}l@{}}0.5729\\   (0.3346)\end{tabular}  & 1.7121  & 0.1377   & 0.3282  \\ \hline
Spain        & \begin{tabular}[c]{@{}l@{}}0.0000\\   (0.5099)\end{tabular} & 0.0000                                             & 0.9999 & \begin{tabular}[c]{@{}l@{}}0.1577\\   (0.5701)\end{tabular}  & 0.2767  & 0.7999   & 0.0248  \\ \hline
Sri Lanka    & \begin{tabular}[c]{@{}l@{}}0.0000\\   (0.243)\end{tabular}  & 0.0000                                             & 0.9999 & \begin{tabular}[c]{@{}l@{}}0.6442\\   (0.2549)\end{tabular}  & 2.5267  & 0.0324** & 0.4150  \\ \hline
Sweden       & \begin{tabular}[c]{@{}l@{}}0.0000\\   (0.2945)\end{tabular} & 0.0000                                             & 0.9999 & \begin{tabular}[c]{@{}l@{}}0.5626\\   (0.3124)\end{tabular}  & 1.8006  & 0.1147   & 0.3165  \\ \hline
Switzerland  & \begin{tabular}[c]{@{}l@{}}0.0000\\   (0.2646)\end{tabular} & 0.0000                                             & 0.9999 & \begin{tabular}[c]{@{}l@{}}0.5537\\   (0.2775)\end{tabular}  & 1.9952  & 0.0771*  & 0.3066  \\ \hline
Thailand     & \begin{tabular}[c]{@{}l@{}}0.0000\\   (0.3558)\end{tabular} & 0.0000                                             & 0.9999 & \begin{tabular}[c]{@{}l@{}}0.3628\\   (0.3804)\end{tabular}  & 0.9538  & 0.3770   & 0.1316  \\ \hline
UK           & \begin{tabular}[c]{@{}l@{}}0.0000\\   (0.3346)\end{tabular} & 0.0000                                             & 0.9999 & \begin{tabular}[c]{@{}l@{}}-0.0688\\   (0.3527)\end{tabular} & -0.1952 & 0.8500   & 0.0047  \\ \hline
USA          & \begin{tabular}[c]{@{}l@{}}0.0000\\   (0.2667)\end{tabular} & 0.0000                                             & 0.9999 & \begin{tabular}[c]{@{}l@{}}0.6063\\   (0.2811)\end{tabular}  & 2.1566  & 0.0631*  & 0.3676 \\ \hline
\end{tabular}
\end{table}


\begin{table}[ht]
\centering
\caption{Regression table: Dependent variable is the scaled centrality and the independent variable is the scaled number of employees (2015-16). *** : significant at 1\%, **: at 5\%, *: at 10\%.}
\begin{tabular}{|l|c|c|c|c|c|c|c|}
\hline
\textbf{Countries} & \textbf{$\beta_0$}                                               & \textbf{Tstat} & \textbf{Pvalue} & \textbf{$\beta_1$}                                               & \textbf{Tstat} & \textbf{Pvalue} & \textbf{Rsquare} \\ \hline
Australia    & \begin{tabular}[c]{@{}l@{}}0.0000\\   (0.3038)\end{tabular} & 0.0000                                             & 0.9999 & \begin{tabular}[c]{@{}l@{}}0.5225\\   (0.3222)\end{tabular}  & 1.6213  & 0.1489    & 0.2730    \\ \hline
Belgium      & \begin{tabular}[c]{@{}l@{}}0.0000\\   (0.2631)\end{tabular} & 0.0000                                             & 0.9999 & \begin{tabular}[c]{@{}l@{}}0.7247\\   (0.2812)\end{tabular}  & 2.5766  & 0.0419**  & 0.5252    \\ \hline
Canada       & \begin{tabular}[c]{@{}l@{}}0.0000\\   (0.2919)\end{tabular} & 0.0000                                             & 0.9999 & \begin{tabular}[c]{@{}l@{}}0.2652\\   (0.3049)\end{tabular}  & 0.8698  & 0.4048    & 0.0703    \\ \hline
Denmark      & \begin{tabular}[c]{@{}l@{}}0.0000\\   (0.3164)\end{tabular} & 0.0000                                             & 0.9999 & \begin{tabular}[c]{@{}l@{}}0.3316\\   (0.3335)\end{tabular}  & 0.9942  & 0.3492    & 0.1099    \\ \hline
Finland      & \begin{tabular}[c]{@{}l@{}}0.0000\\   (0.2762)\end{tabular} & 0.0000                                             & 0.9999 & \begin{tabular}[c]{@{}l@{}}0.5671\\   (0.2912)\end{tabular}  & 1.9474  & 0.0873*   & 0.3216    \\ \hline
France       & \begin{tabular}[c]{@{}l@{}}0.0000\\   (0.1525)\end{tabular} & 0.0000                                             & 0.9999 & \begin{tabular}[c]{@{}l@{}}0.8905\\   (0.1608)\end{tabular}  & 5.5374  & 0.0005*** & 0.7930    \\ \hline
Germany      & \begin{tabular}[c]{@{}l@{}}0.0000\\   (0.2836)\end{tabular} & 0.0000                                             & 0.9999 & \begin{tabular}[c]{@{}l@{}}0.5336\\   (0.299)\end{tabular}   & 1.7845  & 0.1121    & 0.2847    \\ \hline
Greece       & \begin{tabular}[c]{@{}l@{}}0.0000\\   (0.1481)\end{tabular} & 0.0000                                             & 0.9999 & \begin{tabular}[c]{@{}l@{}}-0.8971\\   (0.1561)\end{tabular} & -5.7451 & 0.0004*** & 0.8049    \\ \hline
Hong Kong    & \begin{tabular}[c]{@{}l@{}}0.0000\\   (0.2471)\end{tabular} & 0.0000                                             & 0.9999 & \begin{tabular}[c]{@{}l@{}}0.6759\\   (0.2605)\end{tabular}  & 2.5943  & 0.0318**  & 0.4569    \\ \hline
India        & \begin{tabular}[c]{@{}l@{}}0.0000\\   (0.3354)\end{tabular} & 0.0000                                             & 0.9999 & \begin{tabular}[c]{@{}l@{}}0.0189\\   (0.3535)\end{tabular}  & 0.0534  & 0.9588   & 0.0003    \\ \hline
Indonesia    & \begin{tabular}[c]{@{}l@{}}0.0000\\   (0.2816)\end{tabular} & 0.0000                                             & 0.9999 & \begin{tabular}[c]{@{}l@{}}0.5428\\   (0.2969)\end{tabular}  & 1.8283  & 0.1049    & 0.2947    \\ \hline
Japan        & \begin{tabular}[c]{@{}l@{}}0.0000\\   (0.292)\end{tabular}  & 0.0000                                             & 0.9999 & \begin{tabular}[c]{@{}l@{}}0.4919\\   (0.3078)\end{tabular}  & 1.5982  & 0.1486    & 0.2420    \\ \hline
Malaysia     & \begin{tabular}[c]{@{}l@{}}0.0000\\   (0.2975)\end{tabular} & 0.0000                                             & 0.9999 & \begin{tabular}[c]{@{}l@{}}0.4615\\   (0.3136)\end{tabular}  & 1.4717  & 0.1792    & 0.2130    \\ \hline
Netherlands  & \begin{tabular}[c]{@{}l@{}}0.0000\\   (0.2279)\end{tabular} & 0.0000                                             & 0.9999 & \begin{tabular}[c]{@{}l@{}}0.7686\\   (0.2417)\end{tabular}  & 3.1794  & 0.0155    & 0.5908   \\ \hline
Norway       & \begin{tabular}[c]{@{}l@{}}0.0000\\   (0.2736)\end{tabular} & 0.0000                                             & 0.9999 & \begin{tabular}[c]{@{}l@{}}0.5784\\   (0.2884)\end{tabular}  & 2.0054  & 0.0798*   & 0.3345    \\ \hline
Philippines  & \begin{tabular}[c]{@{}l@{}}0.0000\\   (0.4688)\end{tabular} & 0.0000                                             & 0.9999 & \begin{tabular}[c]{@{}l@{}}0.4191\\   (0.5241)\end{tabular}  & 0.7996  & 0.4823    & 0.1756    \\ \hline
Portugal     & \begin{tabular}[c]{@{}l@{}}0.0000\\   (0.3207)\end{tabular} & 0.0000                                             & 0.9999 & \begin{tabular}[c]{@{}l@{}}0.5425\\   (0.3429)\end{tabular}  & 1.5820  & 0.1647    & 0.2943    \\ \hline
\end{tabular}
\label{Table:regression_employees}
\end{table}

\begin{table}[ht]
\centering
\begin{tabular}{|l|c|c|c|c|c|c|c|}
\hline
Qatar        &                   \begin{tabular}[c]{@{}l@{}} - \\ -\end{tabular}                                           &                                                    &        &                                                              &         &           &           \\ \hline
Saudi Arabia &             \begin{tabular}[c]{@{}l@{}} - \\ -\end{tabular}                                                 &                                                    &        &                                                              &         &           &           \\ \hline
South Africa & \begin{tabular}[c]{@{}l@{}}0.0000\\   (0.3768)\end{tabular} & 0.0000                                             & 0.9999 & \begin{tabular}[c]{@{}l@{}}0.1614\\   (0.4028)\end{tabular}  & 0.4008  & 0.7024    & 0.0260    \\ \hline
Spain        & \begin{tabular}[c]{@{}l@{}}0.0000\\   (0.483)\end{tabular}  & 0.0000                                             & 0.9999 & \begin{tabular}[c]{@{}l@{}}0.3537\\   (0.5400)\end{tabular}  & 0.6550  & 0.5591    & 0.1251    \\ \hline
Sri Lanka    & \begin{tabular}[c]{@{}l@{}}0.0000\\   (0.1924)\end{tabular} & 0.0000                                             & 0.9999 & \begin{tabular}[c]{@{}l@{}}0.7958\\   (0.2018)\end{tabular}  & 3.9435  & 0.0033*** & 0.6334    \\ \hline
Sweden       & \begin{tabular}[c]{@{}l@{}}0.0000\\   (0.315)\end{tabular}  & 0.0000                                             & 0.9999 & \begin{tabular}[c]{@{}l@{}}0.4673\\   (0.3341)\end{tabular}  & 1.3986  & 0.2046    & 0.2184    \\ \hline
Switzerland  & \begin{tabular}[c]{@{}l@{}}0.0000\\   (0.2719)\end{tabular} & 0.0000                                             & 0.9999 & \begin{tabular}[c]{@{}l@{}}0.5177\\   (0.2851)\end{tabular}  & 1.8154  & 0.1028    & 0.2680    \\ \hline
Thailand     &               \begin{tabular}[c]{@{}l@{}} - \\ -\end{tabular}                                               &                                                    &        &                                                              &         &           &           \\ \hline
UK           & \begin{tabular}[c]{@{}l@{}}0.0000\\   (0.255)\end{tabular}  & 0.0000                                             & 0.9999 & \begin{tabular}[c]{@{}l@{}}0.6495\\   (0.2688)\end{tabular}  & 2.4165  & 0.0420**  & 0.4219    \\ \hline
USA          & \begin{tabular}[c]{@{}l@{}}0.0000\\   (0.2529)\end{tabular} & 0.0000                                             & 0.9999 & \begin{tabular}[c]{@{}l@{}}0.6566\\   (0.2666)\end{tabular}  & 2.4628  & 0.0391**  & 0.4312   \\ \hline
\end{tabular}
\end{table}

\begin{table}[ht]
\caption{Regression table: Dependent variable is the scaled centrality and the independent variable is the scaled market capitalization (2008-09). *** : significant at 1\%, **: at 5\%, *: at 10\%.}
\begin{tabular}{|l|c|c|c|c|c|c|c|}
\hline
\textbf{Countries} & \textbf{$\beta_0$}                                               & \textbf{Tstat} & \textbf{Pvalue} & \textbf{$\beta_1$}                                               & \textbf{Tstat} & \textbf{Pvalue} & \textbf{Rsquare} \\ \hline
Australia    & \begin{tabular}[c]{@{}l@{}}0.0000\\   (0.3507)\end{tabular} & 0.0000                                             & 0.9999 & \begin{tabular}[c]{@{}l@{}}0.1770\\   (0.3719)\end{tabular}  & 0.4759  & 0.6485  & 0.0313  \\ \hline
Belgium      & \begin{tabular}[c]{@{}l@{}}0.0000\\   (0.3791)\end{tabular} & 0.0000                                             & 0.9999 & \begin{tabular}[c]{@{}l@{}}-0.1193\\   (0.4053)\end{tabular} & -0.2943 & 0.7783  & 0.0142  \\ \hline
Canada       & \begin{tabular}[c]{@{}l@{}}0.0000\\   (0.3027)\end{tabular} & 0.0000                                             & 0.9999 & \begin{tabular}[c]{@{}l@{}}0.0000\\   (0.3162)\end{tabular}  & 0.0001  & 0.9998  & 0.0000  \\ \hline
Denmark      & \begin{tabular}[c]{@{}l@{}}0.0000\\   (0.3058)\end{tabular} & 0.0000                                             & 0.9999 & \begin{tabular}[c]{@{}l@{}}0.4103\\   (0.3224)\end{tabular}  & 1.2727  & 0.2388  & 0.1683  \\ \hline
Finland      &                     \begin{tabular}[c]{@{}l@{}} - \\ -\end{tabular}                                         &                                                    &        &                                                              &         &         &         \\ \hline
France       & \begin{tabular}[c]{@{}l@{}}0.0000\\   (0.3074)\end{tabular} & 0.0000                                             & 0.9999 & \begin{tabular}[c]{@{}l@{}}0.3995\\   (0.3241)\end{tabular}  & 1.2329  & 0.2526  & 0.1596  \\ \hline
Germany      & \begin{tabular}[c]{@{}l@{}}0.0000\\   (0.2716)\end{tabular} & 0.0000                                             & 0.9999 & \begin{tabular}[c]{@{}l@{}}0.5865\\   (0.2863)\end{tabular}  & 2.0484  & 0.0746* & 0.3440  \\ \hline
Greece       & \begin{tabular}[c]{@{}l@{}}0.0000\\   (0.3327)\end{tabular} & 0.0000                                             & 0.9999 & \begin{tabular}[c]{@{}l@{}}0.1258\\   (0.3507)\end{tabular}  & 0.3587  & 0.7290  & 0.0158  \\ \hline
Hong Kong    & \begin{tabular}[c]{@{}l@{}}0.0000\\   (0.3173)\end{tabular} & 0.0000                                             & 0.9999 & \begin{tabular}[c]{@{}l@{}}0.3239\\   (0.3344)\end{tabular}  & 0.9684  & 0.3611  & 0.1049  \\ \hline
India        & \begin{tabular}[c]{@{}l@{}}0.0000\\   (0.2756)\end{tabular} & 0.0000                                             & 0.9999 & \begin{tabular}[c]{@{}l@{}}0.5696\\   (0.2905)\end{tabular}  & 1.9605  & 0.0855* & 0.3245  \\ \hline
Indonesia    & \begin{tabular}[c]{@{}l@{}}0.0000\\   (0.2925)\end{tabular} & 0.0000                                             & 0.9999 & \begin{tabular}[c]{@{}l@{}}0.4891\\   (0.3083)\end{tabular}  & 1.5861  & 0.1513  & 0.2392  \\ \hline
Japan        & \begin{tabular}[c]{@{}l@{}}0.0000\\   (0.334)\end{tabular}  & 0.0000                                             & 0.9999 & \begin{tabular}[c]{@{}l@{}}0.0915\\   (0.352)\end{tabular}   & 0.2600  & 0.8013  & 0.0083  \\ \hline
Malaysia     & \begin{tabular}[c]{@{}l@{}}0.0000\\   (0.273)\end{tabular}  & 0.0000                                             & 0.9999 & \begin{tabular}[c]{@{}l@{}}0.5809\\   (0.2877)\end{tabular}  & 2.0188  & 0.0782* & 0.3375  \\ \hline
Netherlands  & \begin{tabular}[c]{@{}l@{}}0.0000\\   (0.3165)\end{tabular} & 0.0000                                             & 0.9999 & \begin{tabular}[c]{@{}l@{}}0.4594\\   (0.3357)\end{tabular}  & 1.3686  & 0.2134  & 0.2110  \\ \hline
Norway       & \begin{tabular}[c]{@{}l@{}}0.0000\\   (0.3016)\end{tabular} & 0.0000                                             & 0.9999 & \begin{tabular}[c]{@{}l@{}}0.4369\\   (0.318)\end{tabular}   & 1.3739  & 0.2067  & 0.1909  \\ \hline
Philippines  & \begin{tabular}[c]{@{}l@{}}0.0000\\   (0.4176)\end{tabular} & 0.0000                                             & 0.9999 & \begin{tabular}[c]{@{}l@{}}0.5880\\   (0.4669)\end{tabular}  & 1.2592  & 0.2970  & 0.3457  \\ \hline
Portugal     & \begin{tabular}[c]{@{}l@{}}0.0000\\   (0.3116)\end{tabular} & 0.0000                                             & 0.9999 & \begin{tabular}[c]{@{}l@{}}0.5779\\   (0.3331)\end{tabular}  & 1.7346  & 0.1334  & 0.3340  \\ \hline
\end{tabular}
\label{Table:regression_markcap_08_09}
\end{table}
\begin{table}[ht]
\centering
\begin{tabular}{|l|c|c|c|c|c|c|c|}
\hline
Qatar        & \begin{tabular}[c]{@{}l@{}}0.0000\\   (0.3914)\end{tabular} & 0.0000                                             & 0.9999 & \begin{tabular}[c]{@{}l@{}}0.5140\\   (0.4288)\end{tabular}  & 1.1987  & 0.2967  & 0.2642  \\ \hline
Saudi Arabia & \begin{tabular}[c]{@{}l@{}}0.0000\\   (0.2714)\end{tabular} & 0.0000                                             & 0.9999 & \begin{tabular}[c]{@{}l@{}}0.2179\\   (0.2817)\end{tabular}  & 0.7734  & 0.4542  & 0.0474  \\ \hline
South Africa & \begin{tabular}[c]{@{}l@{}}0.0000\\   (0.3802)\end{tabular} & 0.0000                                             & 0.9999 & \begin{tabular}[c]{@{}l@{}}-0.0932\\   (0.4064)\end{tabular} & -0.2293 & 0.8262  & 0.0086  \\ \hline
Spain        & \begin{tabular}[c]{@{}l@{}}0.0000\\   (0.4885)\end{tabular} & 0.0000                                             & 0.9999 & \begin{tabular}[c]{@{}l@{}}0.3238\\   (0.5462)\end{tabular}  & 0.5928  & 0.5950  & 0.1048  \\ \hline
Sri Lanka    & \begin{tabular}[c]{@{}l@{}}0.0000\\   (0.301)\end{tabular}  & 0.0000                                             & 0.9999 & \begin{tabular}[c]{@{}l@{}}-0.2309\\   (0.3243)\end{tabular}  & -0.7120& 0.4945  & 0.0533  \\ \hline
Sweden       & \begin{tabular}[c]{@{}l@{}}0.0000\\   (0.3435)\end{tabular} & 0.0000                                             & 0.9999 & \begin{tabular}[c]{@{}l@{}}0.3643\\   (0.2658)\end{tabular}  & 0.7297  & 0.4892  & 0.0706  \\ \hline
Switzerland  & \begin{tabular}[c]{@{}l@{}}0.0000\\   (0.3173)\end{tabular} & 0.0000                                             & 0.9999 & \begin{tabular}[c]{@{}l@{}}-0.0532\\   (0.3328)\end{tabular} & -0.1599 & 0.8764  & 0.0028  \\ \hline
Thailand     & \begin{tabular}[c]{@{}l@{}}0.0000\\   (0.3677)\end{tabular} & 0.0000                                             & 0.9999 & \begin{tabular}[c]{@{}l@{}}0.2691\\   (0.3931)\end{tabular}  & 0.6845  & 0.5191  & 0.0724  \\ \hline
UK           & \begin{tabular}[c]{@{}l@{}}0.0000\\   (0.3346)\end{tabular} & 0.0000                                             & 0.9999 & \begin{tabular}[c]{@{}l@{}}0.0691\\   (0.3527)\end{tabular}  & 0.1959  & 0.8495  & 0.0047  \\ \hline
USA          & \begin{tabular}[c]{@{}l@{}}0.0000\\   (0.3315)\end{tabular} & 0.0000                                             & 0.9999 & \begin{tabular}[c]{@{}l@{}}0.1508\\   (0.3495)\end{tabular}  & 0.4315  & 0.6774  & 0.0227 \\ \hline
\end{tabular}
\end{table}

\begin{table}[ht]
\centering
\caption{Regression table: Dependent variable is the scaled centrality and the independent variable is the scaled revenue (2008-09). *** : significant at 1\%, **: at 5\%, *: at 10\%.}
\begin{tabular}{|l|c|c|c|c|c|c|c|}
\hline
\textbf{Countries} & \textbf{$\beta_0$}                                               & \textbf{Tstat} & \textbf{Pvalue} & \textbf{$\beta_1$}                                               & \textbf{Tstat} & \textbf{Pvalue} & \textbf{Rsquare} \\ \hline
Australia    & \begin{tabular}[c]{@{}l@{}}0.0000\\   (0.335)\end{tabular}  & 0.0000                                             & 0.9999 & \begin{tabular}[c]{@{}l@{}}0.3408\\   (0.3553)\end{tabular}  & 0.9592  & 0.3694   & 0.1161    \\ \hline
Belgium      & \begin{tabular}[c]{@{}l@{}}0.0000\\   (0.3676)\end{tabular} & 0.0000                                             & 0.9999 & \begin{tabular}[c]{@{}l@{}}-0.2708\\   (0.3929)\end{tabular} & -0.6891 & 0.5164   & 0.0733    \\ \hline
Canada       & \begin{tabular}[c]{@{}l@{}}0.0000\\   (0.2869)\end{tabular} & 0.0000                                             & 0.9999 & \begin{tabular}[c]{@{}l@{}}0.3194\\   (0.2996)\end{tabular}  & 1.0658  & 0.3115   & 0.1020    \\ \hline
Denmark      & \begin{tabular}[c]{@{}l@{}}0.0000\\   (0.2908)\end{tabular} & 0.0000                                             & 0.9999 & \begin{tabular}[c]{@{}l@{}}0.4982\\   (0.3065)\end{tabular}  & 1.6254  & 0.1427   & 0.2482    \\ \hline
Finland      &               \begin{tabular}[c]{@{}l@{}} - \\ -\end{tabular}                                               &                                                    &        &                                                              &         &          &           \\ \hline
France       & \begin{tabular}[c]{@{}l@{}}0.0000\\   (0.2457)\end{tabular} & 0.0000                                             & 0.9999 & \begin{tabular}[c]{@{}l@{}}0.6805\\   (0.259)\end{tabular}   & 2.6273  & 0.0303** & 0.4631    \\ \hline
Germany      & \begin{tabular}[c]{@{}l@{}}0.0000\\   (0.2792)\end{tabular} & 0.0000                                             & 0.9999 & \begin{tabular}[c]{@{}l@{}}0.5538\\   (0.2943)\end{tabular}  & 1.8814  & 0.0966*  & 0.3067    \\ \hline
Greece       & \begin{tabular}[c]{@{}l@{}}0.0000\\   (0.3071)\end{tabular} & 0.0000                                             & 0.9999 & \begin{tabular}[c]{@{}l@{}}-0.4015\\   (0.3237)\end{tabular} & -1.2401 & 0.2500   & 0.1612    \\ \hline
Hong Kong    & \begin{tabular}[c]{@{}l@{}}0.0000\\   (0.2956)\end{tabular} & 0.0000                                             & 0.9999 & \begin{tabular}[c]{@{}l@{}}0.4719\\   (0.3116)\end{tabular}  & 1.5142  & 0.1684   & 0.2227   \\ \hline
India        & \begin{tabular}[c]{@{}l@{}}0.0000\\   (0.3022)\end{tabular} & 0.0000                                             & 0.9999 & \begin{tabular}[c]{@{}l@{}}0.4333\\   (0.3186)\end{tabular}  & 1.3601  & 0.2108   & 0.1878    \\ \hline
Indonesia    & \begin{tabular}[c]{@{}l@{}}0.0000\\   (0.2747)\end{tabular} & 0.0000                                             & 0.9999 & \begin{tabular}[c]{@{}l@{}}0.5734\\   (0.2896)\end{tabular}  & 1.9798  & 0.0830*  & 0.3288    \\ \hline
Japan        & \begin{tabular}[c]{@{}l@{}}0.0000\\   (0.3226)\end{tabular} & 0.0000                                             & 0.9999 & \begin{tabular}[c]{@{}l@{}}0.2727\\   (0.3401)\end{tabular}  & 0.8018  & 0.4458   & 0.0743    \\ \hline
Malaysia     & \begin{tabular}[c]{@{}l@{}}0.0000\\   (0.2758)\end{tabular} & 0.0000                                             & 0.9999 & \begin{tabular}[c]{@{}l@{}}0.5687\\   (0.2908)\end{tabular}  & 1.9555  & 0.0862*  & 0.3234    \\ \hline
Netherlands  & \begin{tabular}[c]{@{}l@{}}0.0000\\   (0.3406)\end{tabular} & 0.0000                                             & 0.9999 & \begin{tabular}[c]{@{}l@{}}0.2934\\   (0.3613)\end{tabular}  & 0.8122  & 0.4433   & 0.0861    \\ \hline
Norway       & \begin{tabular}[c]{@{}l@{}}0.0000\\   (0.2482)\end{tabular} & 0.0000                                             & 0.9999 & \begin{tabular}[c]{@{}l@{}}0.6723\\   (0.2617)\end{tabular}  & 2.5688  & 0.0331** & 0.4520    \\ \hline
Philippines  & \begin{tabular}[c]{@{}l@{}}0.0000\\   (0.4791)\end{tabular} & 0.0000                                             & 0.9999 & \begin{tabular}[c]{@{}l@{}}0.3728\\   (0.5357)\end{tabular}  & 0.6959  & 0.5365   & 0.1389    \\ \hline
Portugal     & \begin{tabular}[c]{@{}l@{}}0.0000\\   (0.2975)\end{tabular} & 0.0000                                             & 0.9999 & \begin{tabular}[c]{@{}l@{}}0.6268\\   (0.318)\end{tabular}   & 1.9705  & 0.0962*  & 0.3928    \\ \hline
\end{tabular}
\label{Table:regression_rev_08_09}
\end{table}

\begin{table}[ht]
\centering
\begin{tabular}{|l|c|c|c|c|c|c|c|}
\hline
Qatar        & \begin{tabular}[c]{@{}l@{}}0.0000\\   (0.3915)\end{tabular} & 0.0000                                             & 0.9999 & \begin{tabular}[c]{@{}l@{}}-0.5138\\   (0.4289)\end{tabular} & -1.1979 & 0.2970   & 0.2640    \\ \hline
Saudi Arabia & \begin{tabular}[c]{@{}l@{}}0.0000\\   (0.274)\end{tabular}  & 0.0000                                             & 0.9999 & \begin{tabular}[c]{@{}l@{}}0.1724\\   (0.2843)\end{tabular}  & 0.6064  & 0.5555   & 0.0297    \\ \hline
South Africa & \begin{tabular}[c]{@{}l@{}}0.0000\\   (0.3191)\end{tabular} & 0.0000                                             & 0.9999 & \begin{tabular}[c]{@{}l@{}}0.5492\\   (0.3411)\end{tabular}  & 1.6100  & 0.1585   & 0.3017    \\ \hline
Spain        & \begin{tabular}[c]{@{}l@{}}0.0000\\   (0.367)\end{tabular}  & 0.0000                                             & 0.9999 & \begin{tabular}[c]{@{}l@{}}0.7033\\   (0.4103)\end{tabular}  & 1.7139  & 0.1850   & 0.4947    \\ \hline
Sri Lanka    & \begin{tabular}[c]{@{}l@{}}0.0000\\   (0.3167)\end{tabular} & 0.0000                                             & 0.9999 & \begin{tabular}[c]{@{}l@{}}-0.0857\\   (0.3321)\end{tabular}  & -0.2579  & 0.8022  & 0.0073   \\ \hline
Sweden       & \begin{tabular}[c]{@{}l@{}}0.0000\\   (0.3071)\end{tabular} & 0.0000                                             & 0.9999 & \begin{tabular}[c]{@{}l@{}}0.5071\\   (0.3257)\end{tabular}  & 1.5566  & 0.1634   & 0.2571    \\ \hline
Switzerland  & \begin{tabular}[c]{@{}l@{}}0.0000\\   (0.3153)\end{tabular} & 0.0000                                             & 0.9999 & \begin{tabular}[c]{@{}l@{}}0.1245\\   (0.3307)\end{tabular}  & 0.3765  & 0.7152   & 0.0155    \\ \hline
Thailand     & \begin{tabular}[c]{@{}l@{}}0.0000\\   (0.3748)\end{tabular} & 0.0000                                             & 0.9999 & \begin{tabular}[c]{@{}l@{}}0.1908\\   (0.4007)\end{tabular}  & 0.4761  & 0.6507   & 0.0364    \\ \hline
UK           & \begin{tabular}[c]{@{}l@{}}0.0000\\   (0.3293)\end{tabular} & 0.0000                                             & 0.9999 & \begin{tabular}[c]{@{}l@{}}0.1884\\   (0.3472)\end{tabular}  & 0.5427  & 0.6020   & 0.0355    \\ \hline
USA          & \begin{tabular}[c]{@{}l@{}}0.0000\\   (0.3319)\end{tabular} & 0.0000                                             & 0.9999 & \begin{tabular}[c]{@{}l@{}}0.1431\\   (0.3499)\end{tabular}  & 0.4091  & 0.6931   & 0.0205     \\ \hline
\end{tabular}
\end{table}


\begin{table}[ht]
\centering
\caption{Regression table: Dependent variable is the scaled centrality and the independent variable is the scaled number of employees (2008-09). *** : significant at 1\%, **: at 5\%, *: at 10\%.}
\begin{tabular}{|l|c|c|c|c|c|c|c|}
\hline
\textbf{Countries} & \textbf{$\beta_0$}                                               & \textbf{Tstat} & \textbf{Pvalue} & \textbf{$\beta_1$}                                               & \textbf{Tstat} & \textbf{Pvalue} & \textbf{Rsquare} \\ \hline
Australia    & \begin{tabular}[c]{@{}l@{}}0.0000\\   (0.3369)\end{tabular} & 0.0000                                             & 0.9999 & \begin{tabular}[c]{@{}l@{}}0.3251\\   (0.3574)\end{tabular}  & 0.9095  & 0.3932    & 0.1056    \\ \hline
Belgium      & \begin{tabular}[c]{@{}l@{}}0.0000\\   (0.3802)\end{tabular} & 0.0000                                             & 0.9999 & \begin{tabular}[c]{@{}l@{}}-0.0918\\   (0.4065)\end{tabular} & -0.2259 & 0.8287    & 0.0084    \\ \hline
Canada       & \begin{tabular}[c]{@{}l@{}}0.0000\\   (0.2953)\end{tabular} & 0.0000                                             & 0.9999 & \begin{tabular}[c]{@{}l@{}}0.2206\\   (0.3084)\end{tabular}  & 0.7154  & 0.4907    & 0.0486    \\ \hline
Denmark      & \begin{tabular}[c]{@{}l@{}}0.0000\\   (0.3154)\end{tabular} & 0.0000                                             & 0.9999 & \begin{tabular}[c]{@{}l@{}}0.3400\\   (0.3324)\end{tabular}  & 1.0226  & 0.3364    & 0.1156    \\ \hline
Finland      &             \begin{tabular}[c]{@{}l@{}} - \\ -\end{tabular}                                                 &                                                    &        &                                                              &         &           &           \\ \hline
France       & \begin{tabular}[c]{@{}l@{}}0.0000\\   (0.2558)\end{tabular} & 0.0000                                             & 0.9999 & \begin{tabular}[c]{@{}l@{}}0.6465\\   (0.2697)\end{tabular}  & 2.3971  & 0.0433**  & 0.4180    \\ \hline
Germany      & \begin{tabular}[c]{@{}l@{}}0.0000\\   (0.2916)\end{tabular} & 0.0000                                             & 0.9999 & \begin{tabular}[c]{@{}l@{}}0.4939\\   (0.3074)\end{tabular}  & 1.6066  & 0.1468    & 0.2439    \\ \hline
Greece       & \begin{tabular}[c]{@{}l@{}}0.0000\\   (0.3353)\end{tabular} & 0.0000                                             & 0.9999 & \begin{tabular}[c]{@{}l@{}}-0.0177\\   (0.3534)\end{tabular} & -0.0502 & 0.9611    & 0.0003    \\ \hline
Hong Kong    & \begin{tabular}[c]{@{}l@{}}0.0000\\   (0.2864)\end{tabular} & 0.0000                                             & 0.9999 & \begin{tabular}[c]{@{}l@{}}0.5200\\   (0.3019)\end{tabular}  & 1.7222  & 0.1233    & 0.2704    \\ \hline
India        & \begin{tabular}[c]{@{}l@{}}0.0000\\   (0.3088)\end{tabular} & 0.0000                                             & 0.9999 & \begin{tabular}[c]{@{}l@{}}0.3903\\   (0.3255)\end{tabular}  & 1.1991  & 0.2647    & 0.1523    \\ \hline
Indonesia    & \begin{tabular}[c]{@{}l@{}}0.0000\\   (0.2749)\end{tabular} & 0.0000                                             & 0.9999 & \begin{tabular}[c]{@{}l@{}}0.5726\\   (0.2898)\end{tabular}  & 1.9759  & 0.0835*   & 0.3279    \\ \hline
Japan        & \begin{tabular}[c]{@{}l@{}}0.0000\\   (0.2938)\end{tabular} & 0.0000                                             & 0.9999 & \begin{tabular}[c]{@{}l@{}}0.4823\\   (0.3096)\end{tabular}  & 1.5575  & 0.1579    & 0.2326    \\ \hline
Malaysia     & \begin{tabular}[c]{@{}l@{}}0.0000\\   (0.2835)\end{tabular} & 0.0000                                             & 0.9999 & \begin{tabular}[c]{@{}l@{}}0.5341\\   (0.2988)\end{tabular}  & 1.7870  & 0.1117    & 0.2853    \\ \hline
Netherlands  & \begin{tabular}[c]{@{}l@{}}0.0000\\   (0.3165)\end{tabular} & 0.0000                                             & 0.9999 & \begin{tabular}[c]{@{}l@{}}0.4592\\   (0.3357)\end{tabular}  & 1.3679  & 0.2136    & 0.2109    \\ \hline
Norway       & \begin{tabular}[c]{@{}l@{}}0.0000\\   (0.2467)\end{tabular} & 0.0000                                             & 0.9999 & \begin{tabular}[c]{@{}l@{}}0.6772\\   (0.2601)\end{tabular}  & 2.6034  & 0.0314**  & 0.4586    \\ \hline
Philippines  & \begin{tabular}[c]{@{}l@{}}0.0000\\   (0.4808)\end{tabular} & 0.0000                                             & 0.9999 & \begin{tabular}[c]{@{}l@{}}0.3644\\   (0.5376)\end{tabular}  & 0.6777  & 0.5465    & 0.1327    \\ \hline
Portugal     & \begin{tabular}[c]{@{}l@{}}0.0000\\   (0.329)\end{tabular}  & 0.0000                                             & 0.9999 & \begin{tabular}[c]{@{}l@{}}0.5074\\   (0.3517)\end{tabular}  & 1.4425  & 0.1992    & 0.2575    \\ \hline
\end{tabular}
\label{Table:regression_emp_08_09}
\end{table}

\begin{table}[ht]
\centering
\begin{tabular}{|l|c|c|c|c|c|c|c|}
\hline
Qatar        &           \begin{tabular}[c]{@{}l@{}} - \\ -\end{tabular}        &                                                    &        &                                                              &         &           &           \\ \hline
Saudi Arabia &       \begin{tabular}[c]{@{}l@{}} - \\ -\end{tabular}                                                      &                                                    &        &                                                              &         &           &          \\ \hline
South Africa & \begin{tabular}[c]{@{}l@{}}0.0000\\   (0.364)\end{tabular}  & 0.0000                                             & 0.9999 & \begin{tabular}[c]{@{}l@{}}0.3017\\   (0.3892)\end{tabular}  & 0.7753  & 0.4675    & 0.0910    \\ \hline
Spain        & \begin{tabular}[c]{@{}l@{}}0.0000\\   (0.4049)\end{tabular} & 0.0000                                             & 0.9999 & \begin{tabular}[c]{@{}l@{}}0.6205\\   (0.4527)\end{tabular}  & 1.3707  & 0.2640    & 0.3851    \\ \hline
Sri Lanka    & \begin{tabular}[c]{@{}l@{}}0.0000\\   (0.2997)\end{tabular} & 0.0000                                             & 0.9999 & \begin{tabular}[c]{@{}l@{}}0.3330\\   (0.3143)\end{tabular}  & 1.0594  & 0.3170 & 0.1109   \\ \hline
Sweden       & \begin{tabular}[c]{@{}l@{}}0.0000\\   (0.3213)\end{tabular} & 0.0000                                             & 0.9999 & \begin{tabular}[c]{@{}l@{}}0.4321\\   (0.3408)\end{tabular}  & 1.2679  & 0.2453    & 0.1867    \\ \hline
Switzerland  & \begin{tabular}[c]{@{}l@{}}0.0000\\   (0.3083)\end{tabular} & 0.0000                                             & 0.9999 & \begin{tabular}[c]{@{}l@{}}0.2419\\   (0.3234)\end{tabular}  & 0.7481  & 0.4734    & 0.0585    \\ \hline
Thailand     & \begin{tabular}[c]{@{}l@{}}0.0000\\   (0.3618)\end{tabular} & 0.0000                                             & 0.9999 & \begin{tabular}[c]{@{}l@{}}-0.3195\\   (0.3868)\end{tabular} & -0.8259 & 0.4404    & 0.1020    \\ \hline
UK           & \begin{tabular}[c]{@{}l@{}}0.0000\\   (0.2444)\end{tabular} & 0.0000                                             & 0.9999 & \begin{tabular}[c]{@{}l@{}}0.6847\\   (0.2576)\end{tabular}  & 2.6571  & 0.0289**  & 0.4688    \\ \hline
USA          & \begin{tabular}[c]{@{}l@{}}0.0000\\   (0.2776)\end{tabular} & 0.0000                                             & 0.9999 & \begin{tabular}[c]{@{}l@{}}0.5610\\   (0.2926)\end{tabular}  & 1.9171  & 0.0915*   & 0.3148   \\ \hline
\end{tabular}

\end{table}

\end{document}